\documentclass[aps,preprint,eqsecnum]{revtex4-1}

\usepackage{amsmath,amssymb,amsfonts,dcolumn,color,graphicx,graphics,hyperref}
\usepackage{latexsym,placeins,epsfig,tikz}
\usetikzlibrary{arrows,shapes}
\usepackage{subfigure,rotating,hyperref,bm,array}
\usepackage{cancel}
\usepackage{soul}
\usepackage[normalem]{ulem}
\newcommand{\stkout}[1]{\ifmmode\text{\sout{\ensuremath{#1}}}\else\sout{#1}\fi}

\begin{document}

\title{\Large\bf Shadows of rotating wormholes}

\author{Rajibul Shaikh} 
\email{rajibul.shaikh@tifr.res.in}
\affiliation{Tata Institute of Fundamental Research, Homi Bhabha Road, Colaba, Mumbai 400005, India}
\bigskip

\begin{abstract} 
We study shadows cast by a certain class of rotating wormholes and point out the crucial role of a rotating wormhole throat in the formation of a shadow. Overlooking this crucial role of a wormhole throat has resulted incomplete results in the previous studies on shadows of the same class of rotating wormholes. We explore the dependence of the shadows on the spin of the wormholes. We compare our results with that of the Kerr black hole. With increasing values of the spin, the shapes of the wormhole shadows start deviating considerably from that of the black hole. Such considerable deviation, if detected in future observations, may possibly indicate the presence of a wormhole. In other words, the results obtained here indicate that, through the observations of their shadows, the wormholes which are considered in this work and have reasonable spin, can be distinguished from a black hole.
\end{abstract}


\maketitle

\section{Introduction}
It is commonly believed that the supermassive compact region at our Galactic 
Center and those at the centers of many other galaxies contain supermassive 
black holes. However, direct evidence for the
presence of a black hole requires actual detection of the event
horizon. A number of tests have been proposed to confirm the presence of event
horizons in these black hole candidates
\citep{narayan+08,broderick+09,broderick+15}. The evidence is strong
but, of necessity \citep{abramowicz+02}, not conclusive. Typically, a 
black hole event horizon together with a set of unstable light rings (or 
a photon sphere in the case of a spherically symmetric, static black hole) present 
in the exterior geometry of a black hole, is expected to create a characteristic shadow-like 
image (a dark region over a brighter background) of the
radiation emitted by an accretion flow around the black hole or of the photons emitted by nearby light sources. With the purpose of detecting this shadow at mm wavelengths in the 
image of the supermassive compact object Sagittarius A$^*$ (Sgr A$^*$) 
present at our Galactic Center, as well as in the image of that present at the nucleus of the nearby
galaxy M87, the event horizon telescope (EHT) \citep{EHT1,EHT2,EHT3}, an
Earth-spanning millimeter-wave interferometer, is being constructed
and has begun collecting data.

While the intensity map of an
image depends on the details of the accretion process and 
the emission mechanisms, the boundary of the shadow is
only determined by the spacetime metric itself, since it
corresponds to the apparent shape of the photon capture
orbits as seen by a distant observer. Shadows cast by different black holes have been studied
by several researchers in the past, both theoretically and numerically. The shadow of the Schwarzschild black hole was studied by Synge \citep{synge_1966} and Luminet \citep{luminet_1979}. Bardeen studied the shadow cast by the Kerr black hole \citep{bardeen_1973} (see \cite{chandrasekhar} also). Consequently, the shadow of a Kerr black hole and its different aspects (e.g. measurement of the spin parameter) have been studied by several authors \citep{falcke_2000,takahashi_2004,zakharov_2005b,beckwith_2005,
broderick_2006a,takahashi_2007,hioki_2009,paolis_2011}. The shadows cast by other black holes such as Reissner-Nordstrom black holes \citep{zakharov_2005a,zakharov_2014}, binary black holes \citep{yumoto_2012,shipley_2016}, Kerr-Newman black holes \citep{young_1976,vries_2000,takahashi_2005}, Kerr black hole with scalar hair \citep{cunha_2015}, Kerr-Sen black holes \citep{hioki_2008,dastan_2016}, regular black holes \citep{li_2014,abdujabbarov_2016a,amir_2016,sharif_2016,tsukamoto_2018,saha_2018}, Einstein-Maxwell-dilaton-axion black holes \citep{wei_2013}, Kerr-Taub-NUT black holes \citep{abdujabbarov_2013}, Kerr-Newman-NUT black holes \citep{grenzebach_2014}, braneworld black holes \citep{amarilla_2012}, Kaluza-Klein dilaton black holes \citep{amarilla_2013}, Einstein-dilaton-Gauss-Bonnet black holes \citep{cunha_2017}, Horava-Lifshitz black holes \citep{atamurotov_2013b}, non-Kerr black holes \citep{atamurotov_2013,wang_2017}, higher-dimensional black holes \citep{papnoi_2014,amir_2017,abdujabbarov_2015a} and black holes in the presence of plasma \citep{atamurotov_2015,abdujabbarov_2017}, have been studied in the past. Recently, a coordinate-independent characterization of a black hole shadow was discussed in \cite{abdujabbarov_2015b}. A new method for performing general-relativistic ray-tracing calculations to obtain black hole shadow images from a new parametrization of any axisymmetric black hole metric has been discussed in \cite{younsi_2016}. See \cite{shadow_review} for a recent brief review on shadows.

The presence of a shadow, however, does not by itself prove that a compact object is necessarily a black hole. Other horizonless compact objects such as a hard surface \citep{broderick_2006b}, naked singularities \citep{bambi_2009,ortiz_2015,rajibul_2018}, and nonrotating wormholes \citep{bambi_2013a,ohgami_2015,mustafa_2015,ohgami_2016} can also cast similar shadows. Recently, the question of whether a shadow probes the event horizon geometry has been addressed in \cite{shadow_horizon}. In this work, we study shadows of rotating wormholes of Teo class \citep{teo_1998}. Although shadows cast by this class of wormholes have been studied earlier \citep{nedkova_2013,abdujabbarov_2016b}, the results obtained in these works are incomplete since they overlooked the crucial role of a throat of a rotating wormhole in the shadow formation. In this work, we revisit the problem and obtain the complete and correct shapes of the shadows cast by the rotating Teo wormholes. Other than shadows, several other observational aspects of wormholes such as electromagnetic signatures of accretion disks around wormholes \citep{harko_2008,harko_2009,bambi_2013b}, spin precession of a test gyro moving in a rotating wormhole \citep{chakraborty_2017}, and gravitational lensing \citep{cramer_1995,safonova_2001, nandi_2006,abe_2010,nakajima_2012,tsukamoto_2012,tsukamoto_2017,rajibul_2017,jusufi_2018, jusufi_unpub1,jusufi_unpub2}, have been studied in the past. It is also worth mentioning that, although wormholes in general relativity necessarily violate different energy conditions, it is not so in the case of modified gravity wormholes (see \citep{maeda_2008,lobo_2009,dehghani_2009,kanti_2011, harko_2013,rajibul_2015,kar_2015,bronnikov_2015,bambi_2016,rajibul_2016,menchon_2017,moraes_2017, moradpour_2017,hohmann_unpub} and references therein for some recent examples).

The plan of the paper is as follows. In the next section, we briefly recall the rotating Teo wormhole spacetime. In Sec. \ref{sec:null_geodesics}, we briefly summarize the study of null geodesics in the spacetime of a rotating Teo wormhole. We obtain the apparent shapes of shadows cast by the rotating wormholes in Sec. \ref{sec:shadows}. We conclude in Sec. \ref{sec:conclusion} with a summary of the key results.

\section{The rotating wormhole spacetime}
\label{sec:rotating_wormhole}

We start with the stationary, axisymmetric spacetime metric describing a rotating traversable wormhole of Teo class. The metric is given by \citep{teo_1998}
\begin{equation}
ds^2=-N^2 dt^2+\frac{dr^2}{1-\frac{b}{r}}+r^2K^2
\left[d\theta^2+\sin^2\theta(d\varphi-\omega dt)^2
\right],
\label{eq:teo_wormhole}
\end{equation}
where $-\infty<t<\infty$, and $r_0\leq r<\infty$, $0\leq\theta\leq\pi$ and $0\leq\phi\leq 2\pi$ are spherical coordinates. The functions $N$, $b$, $K$, and $\omega$ depend on $r$ and $\theta$ only, such that it is regular on the symmetry axis $\theta=0,\pi$ \citep{teo_1998}. The spacetime describes two identical, asymptotically flat regions joined together at the throat, $r=r_0=b>0$. The above metric is a rotating generalization of the static Morris-Thorne wormhole \citep{morris_1988}.

To ensure that there are no curvature singularities or event horizons, the analog redshift function $N$ must be nonzero and finite everywhere (from the throat to the spatial infinity). Also, to avoid any curvature singularity at the throat and to satisfy the flare-out condition at the throat, the shape function $b$ must satisfy $\partial_\theta b\vert_{r=r_0}=0$, $\partial_r b\vert_{r=r_0}<1$ and $b\leq r$ \citep{teo_1998}. The metric function $K$ determines the area radius given by $R = rK$ and $\omega$ measures the angular velocity of the wormhole.

The metric functions $N$, $K$, $b$ and $\omega$ can be chosen freely provided the above-mentioned regularity conditions are satisfied, thereby obtaining a particular case of the rotating Teo wormhole. A particular choice of the metric functions frequently used in the literature is given by \citep{harko_2009,bambi_2013b,nedkova_2013,abdujabbarov_2016b}
\begin{eqnarray}
N &=&\exp\left[-\frac{r_0}{r}-\alpha\left(\frac{r_0}{r}\right)^\delta\right], \quad b(r)=r_0\left(\frac{r_0}{r}\right)^\gamma,\nonumber \\
K &=&1, \quad \omega = \frac{2J}{r^3},
\label{eq:metric_choice}
\end{eqnarray}
where $J$ is the angular momentum of the wormhole and $\alpha$, $\delta$, and $\gamma$ are real constants. In the case when $\gamma=0$, we can write $r_0=2M$, $M$ being the mass of the wormhole.

Let us now show that, for $\gamma=0$, $M$ defined above is the ADM mass of the wormholes. For asymptotic flat spacetimes, the ADM mass can be calculated from \citep{wald}
\begin{equation}
m_{ADM}=\frac{1}{16\pi}\lim_{r\to \infty}\sum_{\mu,\nu=1}^3 \int_{S} (\partial_\mu h_{\mu\nu}-\partial_\nu h_{\mu\mu}) N^\nu dS,
\label{eq:ADM_expression}
\end{equation}
where $r=\sqrt{(x^1)^2+(x^2)^2+(x^3)^2}$, ($x^1$, $x^2$, $x^3$) are asymptotically Euclidean coordinates for the spacelike hypersurface $\Sigma$ of constant $t$, $h_{\mu\nu}$ is the induced metric on  $\Sigma$, $S$ is a topological two-sphere of constant $r$ (in $\Sigma$) with outward pointing unit normal $N^\nu$ and $dS$ is an area element. For the wormholes, the metric on $\Sigma$ is given by
\begin{equation}
ds_\Sigma^2=\frac{dr^2}{1-\frac{b}{r}}+r^2K^2
\left[d\theta^2+\sin^2\theta d\varphi^2\right],
\end{equation}
where we take $K=K(r)$. However, calculating the integral in (\ref{eq:ADM_expression}) is a bit tricky since one has to switch over to the pseudo-Cartesian coordinate system ($x^1$, $x^2$, $x^3$) associated to the spherical coordinate system $(r,\theta,\varphi)$. According to \cite{ADM_note} [see Eqs. (1.1.32)--(1.1.37) of \cite{ADM_note} where $x^1=x$, $x^2=y$ and $x^3=z$], for the metric
\begin{equation}
ds_\Sigma^2=\phi(r) dr^2+r^2\chi(r)
\left[d\theta^2+\sin^2\theta d\varphi^2\right]
\end{equation}
on $\Sigma$, the expression (\ref{eq:ADM_expression}) for the ADM mass reduces to
\begin{equation}
m_{ADM}=\lim_{r\to \infty}\frac{1}{2}\left[-r^2\chi'+r(\phi-\chi)\right].
\end{equation}
In our case, for the metric choice in (\ref{eq:metric_choice}),
\begin{equation}
\phi(r)=\frac{1}{1-\frac{r_0}{r}}, \quad \chi(r)=K^2=1,
\end{equation}
where we have taken $\gamma=0$, as is considered in the rest of the paper. Therefore, we can immediately find that the ADM mass for the wormholes is given by
\begin{equation}
m_{ADM}=\frac{r_0}{2},
\end{equation}
which implies that $m_{ADM}=M$ for $\gamma=0$. Therefore, in the rest of the paper, we take $r_0=2M$ for $\gamma=0$. Here, we would like to point out that the wormholes in Eq. (\ref{eq:teo_wormhole}) are symmetric wormholes since the two regions, which are connected through the throat, are represented by the same copy of the spacetime geometry. For such symmetric wormholes, $M$ mentioned above is the ADM mass as seen by an observer located at the asymptotic spatial infinity of any of the two regions.

The rotating wormhole \ref{eq:teo_wormhole} may contain an ergoregion around its throat. An ergoregion is a region where $g_{tt}=-(N^2-\omega^2 r^2 K^2\sin^2\theta)\geq 0$. The boundary of the ergoregion, i.e., the ergosurface is given by $g_{tt}=0$. For the metric (\ref{eq:teo_wormhole}), the ergosurface is given by
\begin{equation}
N^2-\omega^2 r^2 K^2\sin^2\theta=0.
\end{equation}
\begin{figure}[ht]
\centering
\subfigure[$J/M^2=0.9$]{\includegraphics[scale=0.45]{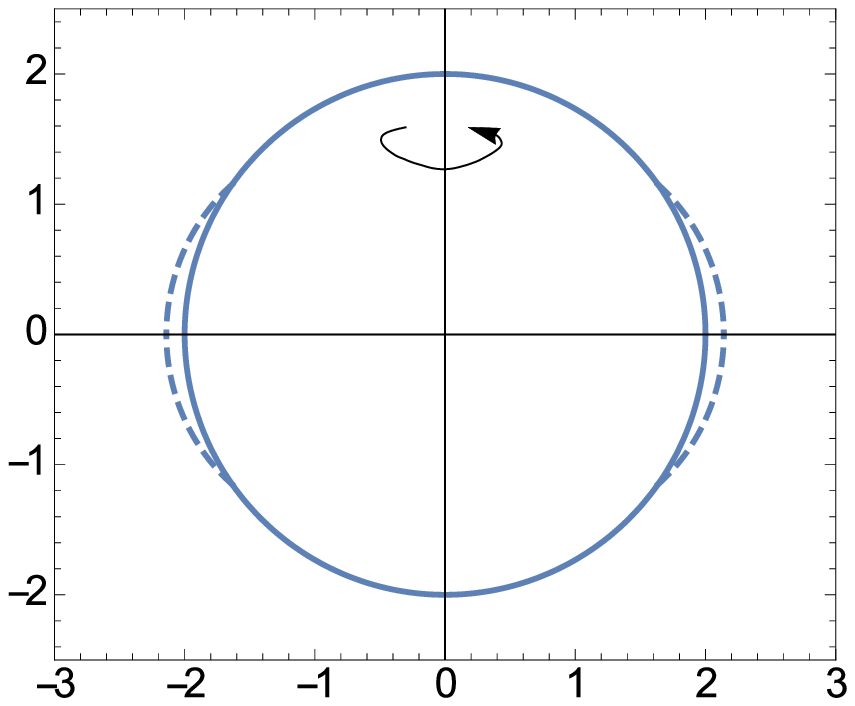}}
\subfigure[$J/M^2=1.2$]{\includegraphics[scale=0.45]{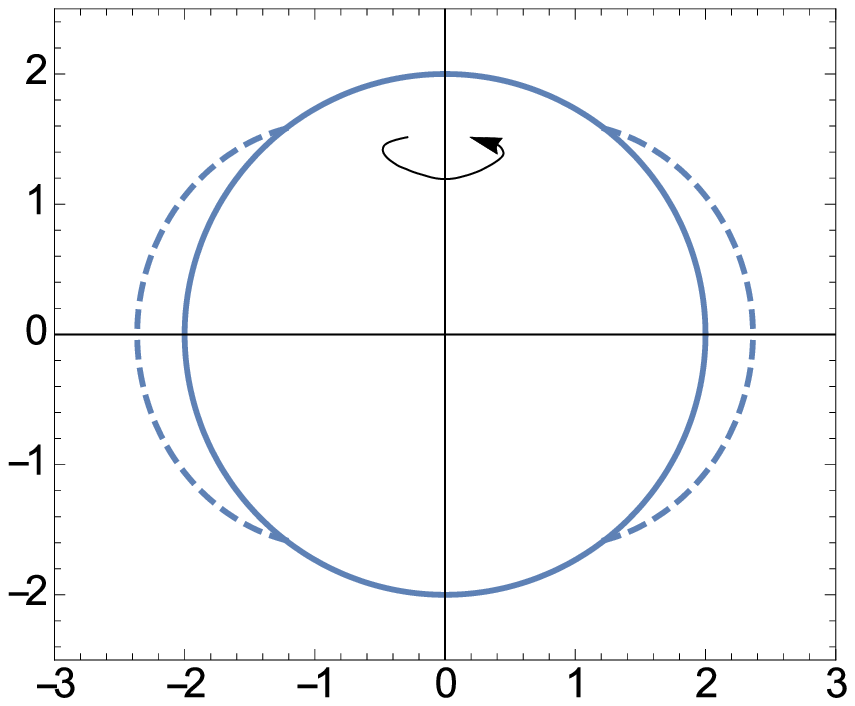}}
\subfigure[$J/M^2=1.5$]{\includegraphics[scale=0.45]{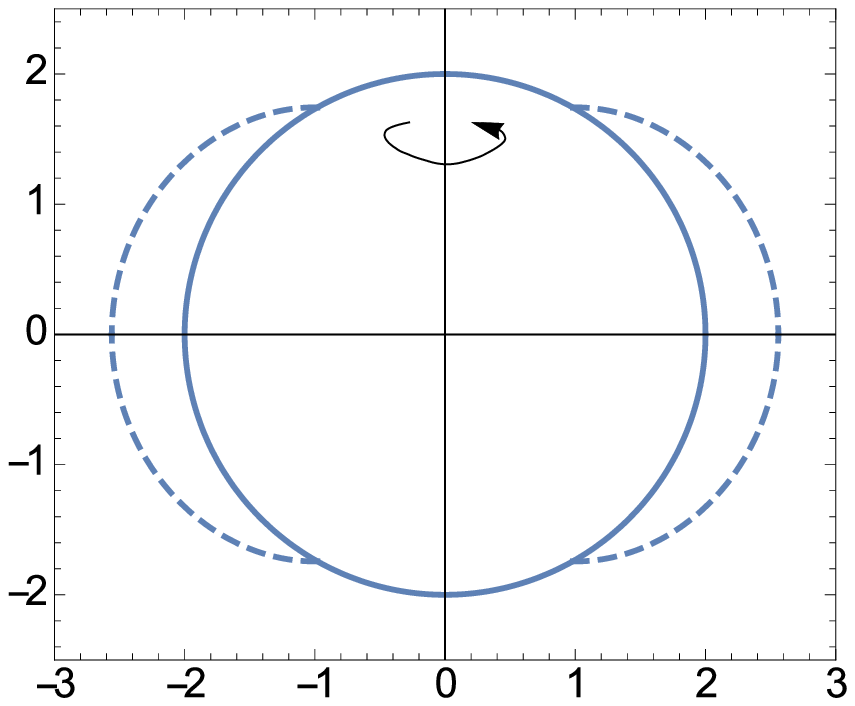}}
\subfigure[$J/M^2=2.0$]{\includegraphics[scale=0.45]{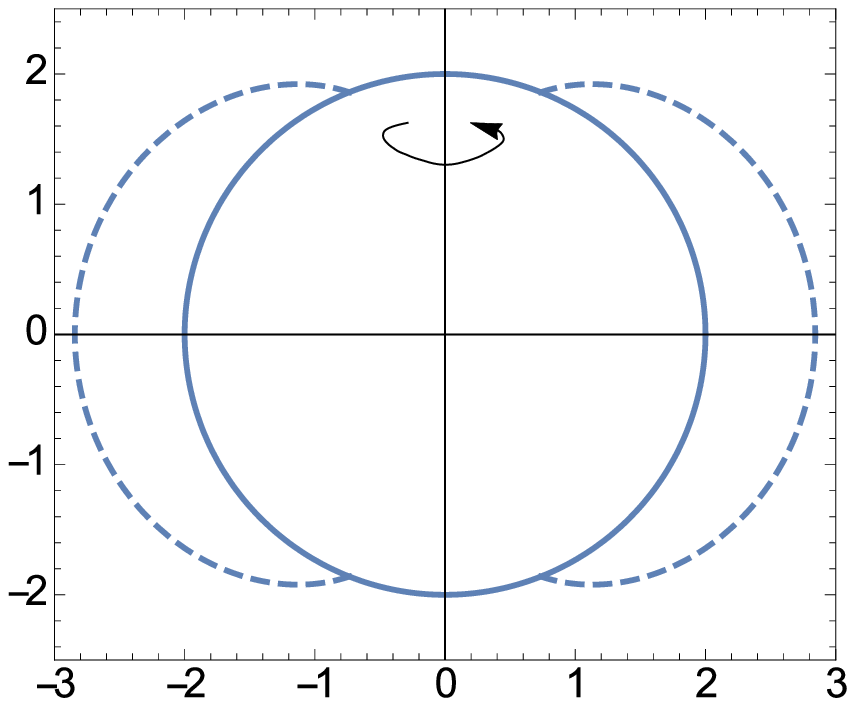}}
\caption{The ergoregion of a rotating wormhole (in units of $M$) for the metric functions 
   (\ref{eq:metric_choice}) with $\alpha=0$, $\gamma=0$ and $\delta=0$. The region 
   between the solid curve (wormhole throat) and the dashed curve (ergosurface) is the 
   ergoregion. The arrow indicates the spin axis of the wormhole.}
\label{fig:ergoregion}
\end{figure}
Figure \ref{fig:ergoregion} shows the ergoregion of a rotating wormhole for the choice of metric functions (\ref{eq:metric_choice}) with $\alpha=0$, $\gamma=0$ and $\delta=0$. Note that the ergoregion does not extend up to the poles $\theta=0$ and $\theta=\pi/2$. It exists between angles $\theta_c$ and $(\pi-\theta_c)$ (i.e., it touches the throat at these angles), where $0<\theta_c\leq \pi/2$ and
\begin{equation}
\sin\theta_c=\Big\vert \frac{N_0}{\omega_0 r_0 K_0} \Big\vert.
\end{equation}
In the above, the subscript `0' implies that the metric functions are evaluated at the throat. Also, note that the ergoregion exists when the spin $a$ ($=J/M^2$) crosses a certain critical value $a_c$ given by $\sin\theta_c= 1$, i.e., by $\omega_c=\frac{N_0}{r_0 K_0}$. For the metric functions used in Fig. \ref{fig:ergoregion}, the critical spin is given by $a_c=J_c/M^2\simeq 0.74$.

\section{Null geodesics in a rotating wormhole spacetime}
\label{sec:null_geodesics}

Null geodesics in the spacetime of the wormholes (\ref{eq:teo_wormhole}) have already been studied in \citep{nedkova_2013}. However, we summarize them here for the calculation purpose of shadows obtained in the next section. The Lagrangian describing the motion of a photon in the spacetime of the rotating wormhole (\ref{eq:teo_wormhole}) is given by
\begin{equation}
2\mathcal{L}=-N^2 \dot{t}^2+\frac{\dot{r}^2}{1-\frac{b}{r}}+r^2K^2
\left[\dot{\theta}^2+\sin^2\theta(\dot{\varphi}-\omega \dot{t})^2
\right],
\end{equation}
where an overdot represents a differentiation with respect to the affine parameter $\lambda$. Since the Lagrangian is independent of $t$ and $\phi$, we have two constants of motion, namely, the energy $E$ and the angular momentum $L$ (about the axis of symmetry) of the photon:
\begin{equation}
p_t=\frac{\partial \mathcal{L}}{\partial \dot{t}}=-N^2 \dot{t}-\omega r^2K^2 \sin^2\theta(\dot{\varphi}-\omega \dot{t})=-E
\end{equation}
\begin{equation}
p_\phi=\frac{\partial \mathcal{L}}{\partial \dot{\phi}}=r^2K^2 \sin^2\theta(\dot{\varphi}-\omega \dot{t})=L.
\end{equation}
Solving the last two equations, we obtain
\begin{equation}
\dot{t}=\frac{E-\omega L}{N^2}, \quad \dot{\varphi}=\frac{L}{r^2K^2\sin^2\theta}+\frac{\omega(E-\omega L)}{N^2}.
\end{equation}
The $r$- and $\theta$-component of the momentum are, respectively, given by
\begin{equation}
p_r=\frac{\partial \mathcal{L}}{\partial \dot{r}}=\frac{\dot{r}}{1-\frac{b}{r}}, \quad p_\theta=\frac{\partial \mathcal{L}}{\partial \dot{\theta}}=r^2K^2
\dot{\theta}.
\label{eq:r_momentum}
\end{equation}
The $r$- and $\theta$-part of the geodesic equations can be obtained by solving the Hamilton-Jacobi equation
\begin{equation}
\frac{\partial S}{\partial \lambda}=-\frac{1}{2}g^{\mu\nu}\frac{\partial S}{\partial x^{\mu}}\frac{\partial S}{\partial x^{\nu}},
\label{eq:HJ}
\end{equation}
where $S$ is the Jacobi action. If there is a separable solution, then in terms of the already known constants of the motion, it must take the form
\begin{equation}
S=\frac{1}{2}\mu ^2 \lambda - E t + L \varphi + S_{r}(r)+S_{\theta}(\theta),
\label{eq:action_ansatz}
\end{equation}
where $\mu $ is the mass of the test particle. For photons, $\mu=0$. If the metric functions $N$, $b$, $K$, and $\omega$ of the rotating wormhole (\ref{eq:teo_wormhole}) are functions of the radial coordinates $r$ only, then the Hamilton-Jacobi equation is separable. Inserting Eq. (\ref{eq:action_ansatz}) into Eq. (\ref{eq:HJ}) and separating out the $r$- and $\theta$-part, we obtain \citep{nedkova_2013}
\begin{equation}
\left(\frac{dS_\theta}{d\theta}\right)^2=Q-\frac{L^2}{\sin^2\theta},
\label{eq:HJ_theta}
\end{equation}
\begin{equation}
\left(1-\frac{b}{r}\right)N^2\left(\frac{dS_r}{dr}\right)^2=\left(E-\omega L\right)^2-\mu^2N^2- Q \frac{N^2}{r^2K^2},
\label{eq:HJ_r}
\end{equation}
where $Q$ is the Carter constant. Since $p_r=\frac{\partial S}{\partial r}=\frac{dS_r}{dr}$ and $p_\theta=\frac{\partial S}{\partial \theta}=\frac{dS_\theta}{d\theta}$, using Eqs. (\ref{eq:r_momentum}), (\ref{eq:HJ_theta}), and (\ref{eq:HJ_r}), we obtain \citep{nedkova_2013}
\begin{equation}
\frac{N}{\left(1-\frac{b}{r}\right)^{1/2}}\frac{dr}{d\lambda} = \pm\sqrt{R(r)}, \quad
r^2K^2\frac{d\theta}{d\lambda} = \pm\sqrt{T(\theta)},
\end{equation}
where
\begin{equation}
R(r)=\left(E-\omega L\right)^2 - \mu^2N^2 - Q \frac{N^2}{r^2K^2},
\end{equation}
\begin{equation}
T(\theta) = Q - \frac{L^2}{\sin^2\theta}.
\end{equation}
Although there are three constants of motion $E$, $L$, and $Q$, the geodesic motion of a photon is characterized by two independent parameters defined by \citep{nedkova_2013}
\begin{equation}
\xi=\frac{L}{E},\quad \eta=\frac{Q}{E^2}.
\end{equation}
$\xi$ and $\eta$ are known as impact parameters. By introducing a new affine parameter $\tilde{\lambda}= E\lambda$, we can redefine the functions $R(r)$ and $T(\theta)$ as
\begin{equation}
R(r)= \left(1-\omega \xi\right)^2 - \eta \frac{N^2}{r^2K^2}, \quad T(\theta)= \eta - \frac{\xi^2}{\sin^2\theta},
\end{equation}
where we have taken $\mu=0$ for a photon.

For latter use, we write down the radial equation of motion in the following form:
\begin{equation}
\left(\frac{dr}{d\tilde{\lambda}}\right)^2 + V_{eff} = 0, \quad V_{eff} =- \frac{1}{N^2}\left(1-\frac{b}{r}\right)R(r),
\label{eq:effective_pot}
\end{equation}
where $V_{eff}$ is the effective potential describing the geodesic motion of a photon.

\section{Shadows of rotating wormholes}
\label{sec:shadows}

A wormhole connects two regions of spacetime. To obtain the shadow cast by a wormhole, we assume that the wormhole is illuminated by a light source situated in one of the regions, and no light sources are present in the vicinity of the throat in the other region \citep{nedkova_2013}. Depending on the impact parameters, some of the photons from the light source plunge into the wormhole, pass through the throat and go to the other region, and some get scattered away to the infinity of the first region. A distant observer situated in the first region only receives the scattered photons. Therefore, in the observer's sky, the scattered photons form bright spots, whereas the photons captured by the wormhole form dark spots. The union of the dark spots in the observer's sky constitutes the shadow.

Therefore, the task is to find out the critical orbits that separate the escaping and plunging photons. These critical orbits are characterized by certain critical values of the impact parameters $\xi$ and $\eta$. A small perturbation in these critical values can turn it either to an escape or to a capture orbit. Therefore, the critical impact parameters define the boundary of a shadow. The critical orbits are unstable circular photon orbits corresponding to the highest maximum of the effective potential $V_{eff}$. The unstable circular photon orbits are determined by the standard conditions
\begin{eqnarray}
V_{eff} = 0,\quad \frac{dV_{eff}}{dr} = 0,\quad \frac{d^2 V_{eff}}{dr^2} \leq 0.
\label{eq:pot_maxima}
\end{eqnarray}
Note that, while writing the above set of conditions in terms of $R(r)$, one must be careful about the throat where $(1-\frac{b}{r})=0$ [see Eq. (\ref{eq:effective_pot})]. For the unstable circular orbits whose radii do not coincide with the throat radius $r_0$ [i.e., for the unstable circular orbits lying outside the throat such that $(1-\frac{b}{r})\neq 0$ on the circular orbits], in terms of $R(r)$, the above set of conditions can be written as
\begin{eqnarray}
R(r) = 0,\quad \frac{dR}{dr}=0,\quad \frac{d^2R}{dr^2}\geq 0.
\label{eq:R_maxima}
\end{eqnarray}
Using $R=0$ and $dR/dr=0$, we obtain \citep{nedkova_2013}
\begin{equation}
\eta=\left[\frac{r^2K^2}{N^2}(1-\omega\xi)^2\right]_{r=r_{ph}},
\label{eq:eta_ph}
\end{equation}
\begin{equation}
\xi=\frac{\Sigma}{\Sigma\omega-\omega'}\Big\vert_{r=r_{ph}}, \quad \Sigma=\frac{1}{2}\frac{d}{dr}\ln\left(\frac{N^2}{r^2K^2}\right),
\label{eq:xi_ph}
\end{equation}
where $r_{ph}$ is the radius of a circular photon orbit. The authors in \cite{nedkova_2013} used the above set of equations for $\xi$ and $\eta$ to obtain the shadows. But, they overlooked the fact that the effective potential can exhibit an extremum at the wormhole throat, thereby implying circular photon orbits (either stable or unstable) at the throat. For example, for a nonrotating wormhole, the wormhole throat acts as the position of the maximum of the potential when there are no extrema outside the throat, the throat thus being the position of unstable circular orbits and hence deciding the boundary of a shadow \citep{bambi_2013a,ohgami_2015}. In our case, since $(1-\frac{b}{r})$ vanishes at the throat $r=r_0$, it can be seen from (\ref{eq:effective_pot}) that the effective potential has a maximum, i.e., Eq. (\ref{eq:pot_maxima}) is satisfied at the throat when
\begin{equation}
R(r_0)=0, \quad \frac{d^2R}{dr^2}\Big\vert_{r=r_0}\geq 0,
\end{equation}
which is different from (\ref{eq:R_maxima}). For circular photon orbits at the throat, we have $R(r_0)=0$, i.e.,
\begin{equation}
\left(1-\omega_0 \xi\right)^2 - \eta \frac{N_0^2}{r_0^2K_0^2}=0,
\label{eq:xieta_throat}
\end{equation}
where the subscript `0' implies that the functions are evaluated at the throat.

For given metric functions and metric parameter values, the parametric plots of (\ref{eq:eta_ph}) and (\ref{eq:xi_ph}) with respect to the parameter $r_{ph}$ plus the plot of (\ref{eq:xieta_throat}) define the boundary of a shadow in the impact parameter space. However, in realistic observation, the apparent shape of a shadow is measured in the observer's sky, a plane passing through the center of the wormhole and normal to the line of sight joining the observer and the center of the wormhole. The coordinates on this plane are denoted by $\alpha$ and $\beta$ and are known as celestial coordinates.

In terms of the tangent to a photon geodesic at the observer's position, the celestial coordinates are given as \cite{vazquez_2004}
\begin{equation}
\alpha=\lim_{r\rightarrow \infty}\left( -r^{2}\sin\theta_{obs}\frac{d\varphi}{dr}\right), \quad
\beta=\lim_{r\rightarrow \infty}r^{2}\frac{d\theta}{dr},
\label{eq:alpha}
\end{equation}
where $\theta_{obs}$ is the inclination angle, i.e., the angle between the rotation axis of the wormhole and the line of sight of the observer. Using the geodesic equations, we obtain \citep{nedkova_2013}
\begin{equation}
\alpha= -\frac{\xi}{\sin\theta_{obs}}, \quad
\beta= \left(\eta - \frac{\xi^2}{\sin^2\theta_{obs}}\right)^{1/2}.
\label{eq:celestial}
\end{equation}
In the $(\alpha,\beta)$-plane, Eqs. (\ref{eq:eta_ph}), (\ref{eq:xi_ph}) and (\ref{eq:celestial}) define the part of the boundary of the shadow formed by the unstable circular photon orbits which lie outside the throat. However, for the part of the boundary of the shadow which is due to the unstable circular orbits at the throats, we have, from Eqs. (\ref{eq:xieta_throat}) and (\ref{eq:celestial}),
\begin{equation}
(N_0^2-\omega_0^2 r_0^2K_0^2\sin^2\theta_{obs})\alpha^2-2\omega_0 r_0^2K_0^2\sin\theta_{obs}\alpha-r_0^2K_0^2+N_0^2\beta^2=0.
\label{eq:alphabeta_throat}
\end{equation}
The authors in \cite{nedkova_2013} overlooked the above-mentioned contribution of the throat in the shadow formation. Therefore, the results they obtained are incomplete. The correct and complete results are obtained when one considers the last equation as well. To show it explicitly, we consider the following metric functions:
\begin{equation}
N =\exp\left[-\frac{r_0}{r}\right], \quad b(r)=r_0=2M,\quad K=1, \quad \omega = \frac{2J}{r^3}.
\label{eq:metric_choice_1}
\end{equation}
\begin{figure}[ht]
\centering
\subfigure[]{\includegraphics[scale=0.47]{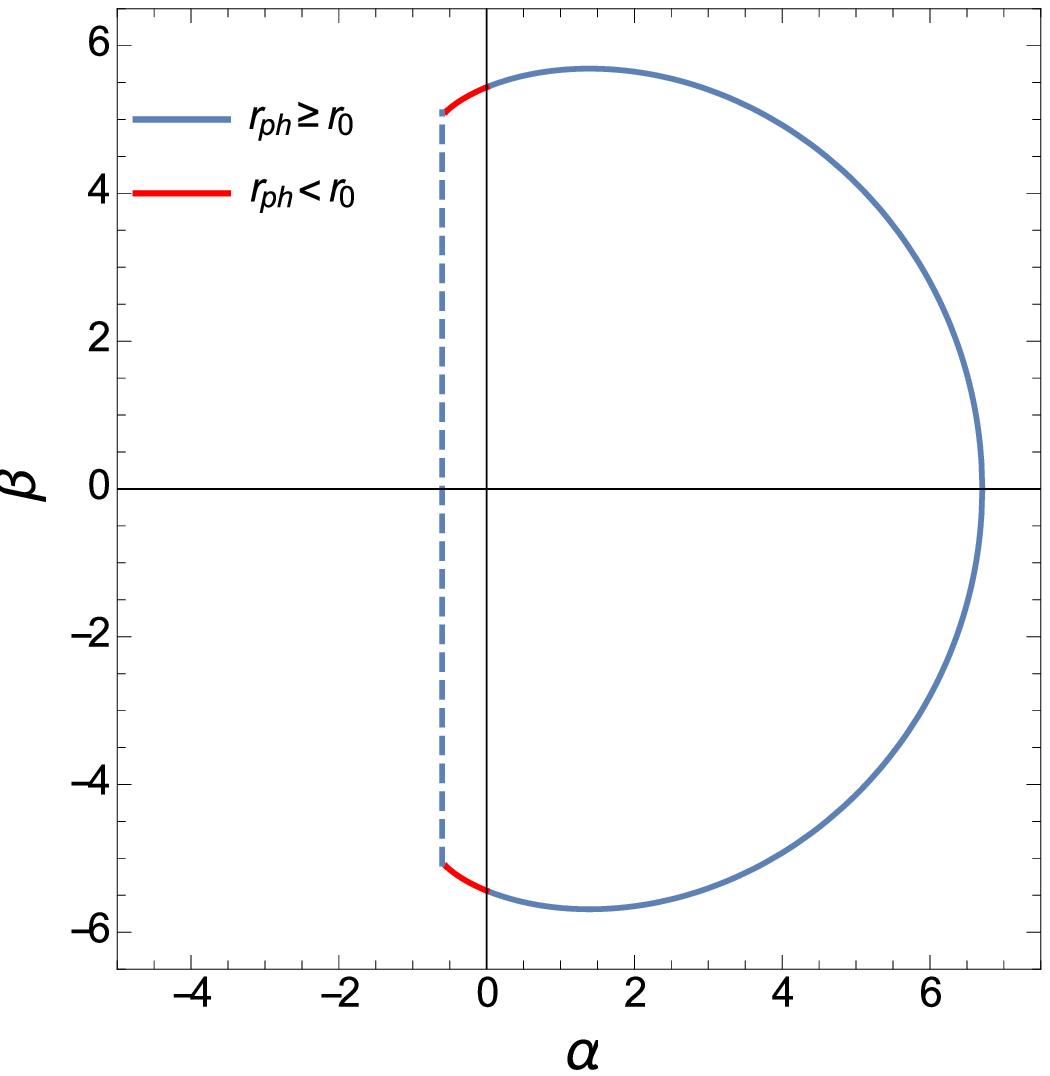}\label{fig:ows_general1}}
\subfigure[]{\includegraphics[scale=0.55]{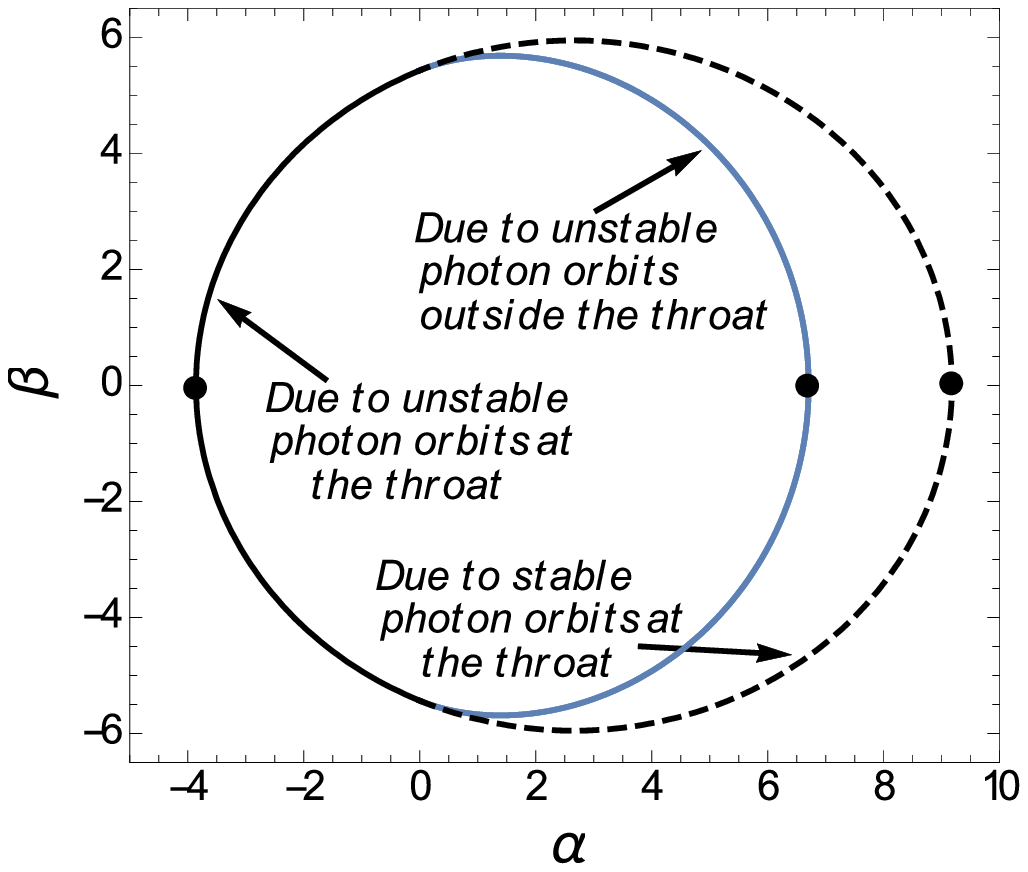}\label{fig:ows_genera2}}
\subfigure[]{\includegraphics[scale=0.47]{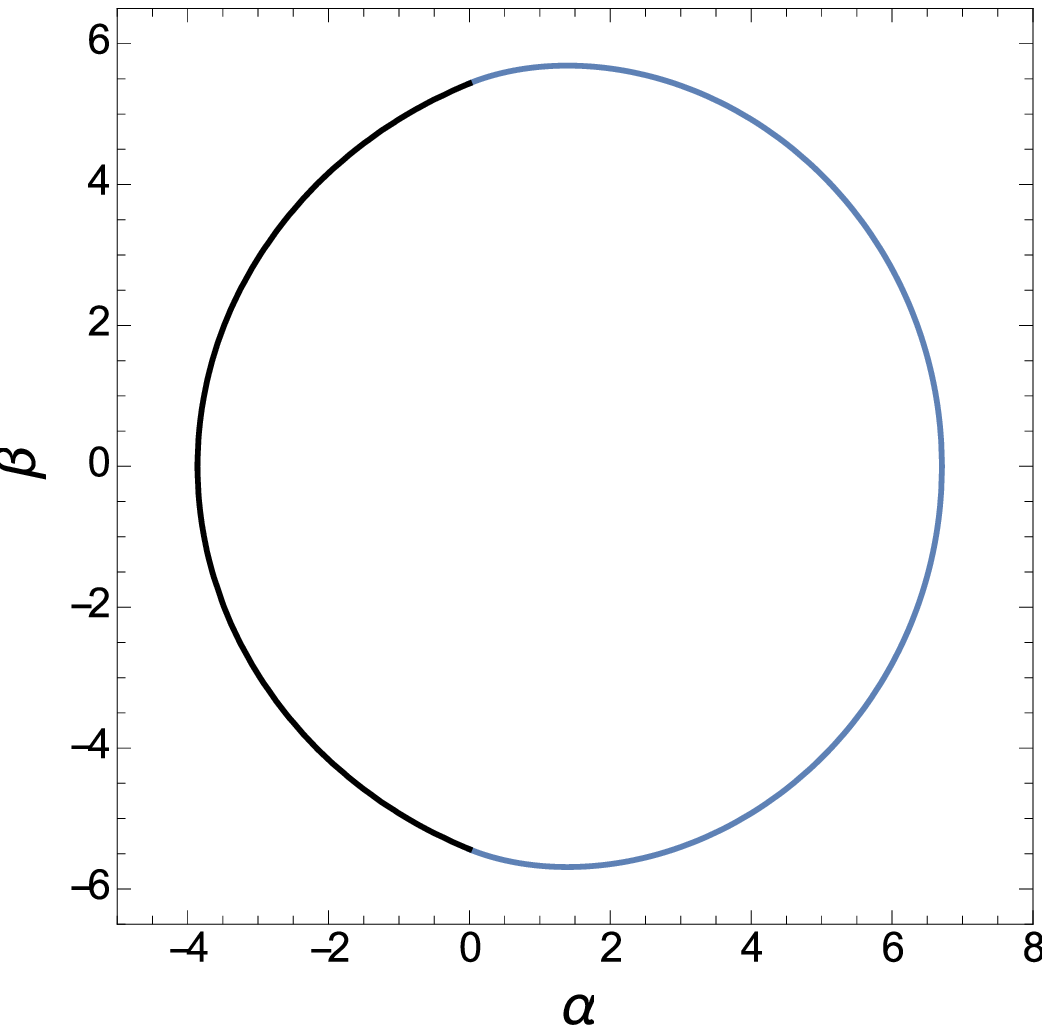}\label{fig:ows_general3}}
\caption{Plots showing how the shape of a shadow looks like when the contribution of the throat 
   [Eq. (\ref{eq:alphabeta_throat})] (a) is not and (b) is considered. Here, the metric functions are given by (\ref{eq:metric_choice_1}) and $J/M^2$=0.3. Panel (c) shows the actual shape of the shadow. The vertical dashed line in (a) is drawn by hand. The impact parameters denoted by the black dots are used in Fig. \ref{fig:ows_pot1}.}
\label{fig:ows_general}
\end{figure}
In Fig. \ref{fig:ows_general}, we have plotted an example of a shadow for the metric parameter values corresponding to the third figure in the first row of Fig. 1 of \cite{nedkova_2013}. The corresponding shadow obtained without considering the contribution of the throat, i.e., without considering Eq. \ref{eq:alphabeta_throat} (which is the case in \cite{nedkova_2013}) is shown in Fig. \ref{fig:ows_general1}. When the contribution of the throat is taken into consideration, we obtain Fig. \ref{fig:ows_genera2}. The correct shape of the shadow shown in Fig. \ref{fig:ows_general3} is given by the common region enclosed by the curve obtained from Eqs. (\ref{eq:eta_ph}), (\ref{eq:xi_ph}), and (\ref{eq:celestial}) [the blue curve in Fig. \ref{fig:ows_genera2}] and the curve obtained from Eq. (\ref{eq:alphabeta_throat}) [the black curve in Fig. \ref{fig:ows_genera2}]. Moreover, from the red part of the plot in Fig. \ref{fig:ows_general1}, it seems that, while plotting the parametric plot using (\ref{eq:eta_ph}), (\ref{eq:xi_ph}), and (\ref{eq:celestial}), the authors in \cite{nedkova_2013} have considered $r_{ph}<r_0$ also, which is not correct since the wormhole spacetime is valid only for $r\geq r_0$.

The reason why the shadow is given by the common region enclosed by the curves mentioned above, can be understood as follows. A circular photon orbit (stable or unstable) with given critical impact parameters lies in a plane which passes through the center of the wormhole and makes an angle $\theta_{ph}$ with the rotation axis, where $T(\theta_{ph})=0$. If, in a given plane, there are no stable and unstable photon orbits outside the throat, then the throat acts as a position of the unstable photon orbit, i.e., the potential has a maximum at the throat. The solid black curve in Fig. \ref{fig:ows_genera2} corresponds to this situation. However, in a given plane, if there are stable or unstable photon orbits outside the throat, then the throat also acts as a position of either stable or unstable photon orbits. Figure \ref{fig:ows_genera2} shows that, in the case where there is an unstable photon orbit (solid blue curve) outside the throat, the throat acts as the position of a stable photon orbit (dashed black curve). To elaborate it further, we choose different values of $\alpha$ and $\beta$ (which is the equivalent of choosing $\xi$ and $\eta$) and plot the effective potential. However, to show the effective potential for both the regions of the wormhole, we use the proper radial distance given by \citep{morris_1988}
\begin{equation}
l(r)=\pm\int_{r_0}^r\frac{dr}{\sqrt{1-\frac{b}{r}}},
\end{equation}
where the plus sign corresponds to one region and the minus sign to the other region. The throat is at $l(r_0)=0$.

\begin{figure}[ht]
\centering
\subfigure[]{\includegraphics[scale=0.55]{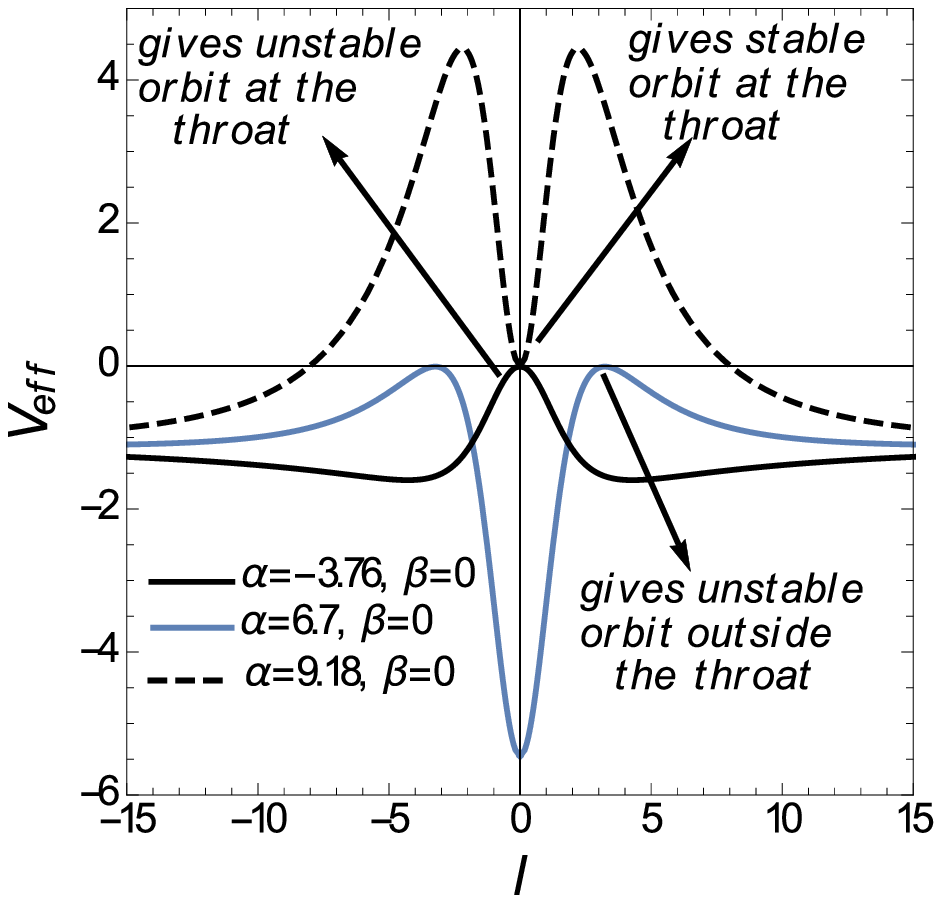}\label{fig:ows_pot1}}
\subfigure[]{\includegraphics[scale=0.55]{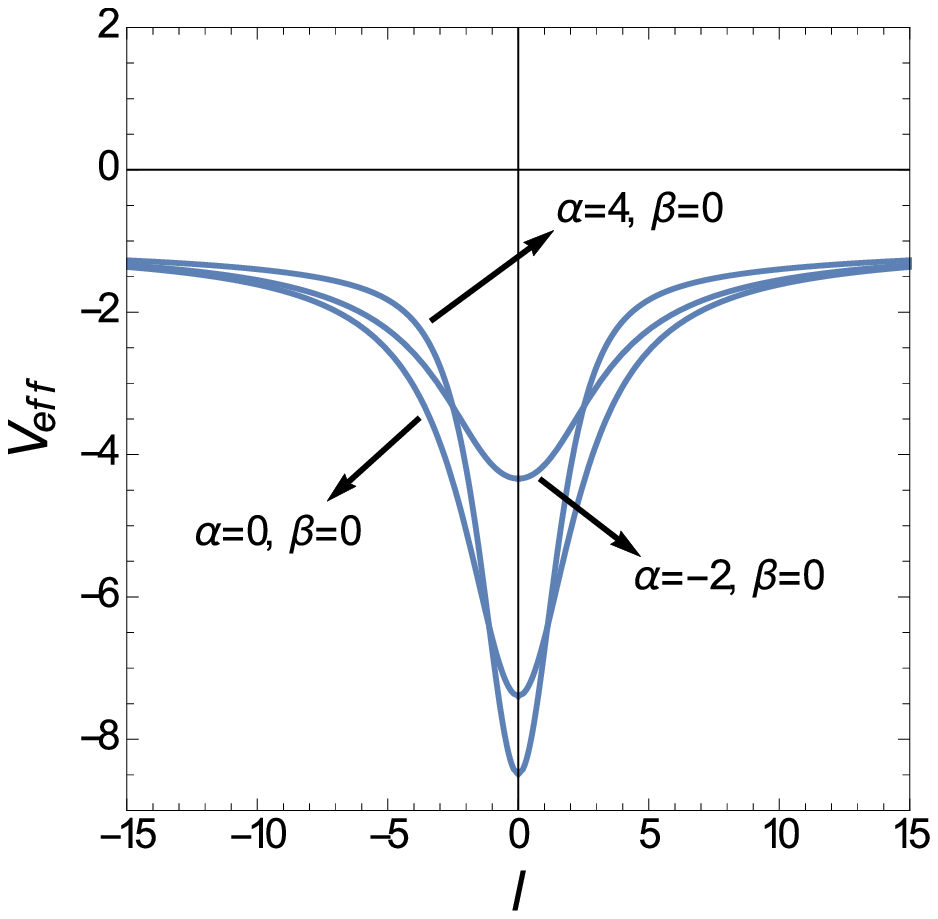}\label{fig:ows_pot2}}
\subfigure[]{\includegraphics[scale=0.55]{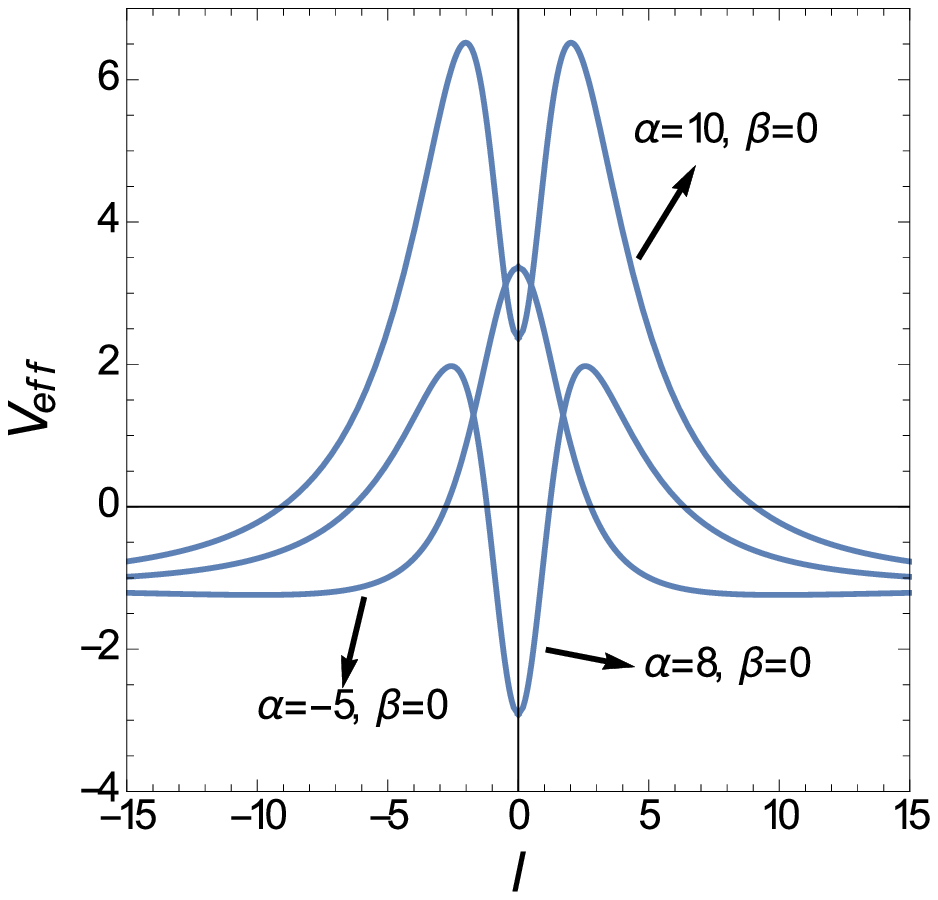}\label{fig:ows_pot3}}
\caption{Plots of the effective potential for (a) impact parameter values denoted by the black dots of Fig. \ref{fig:ows_genera2} and for those lying (b) inside and (c) outside of the shadow boundary shown in Fig. \ref{fig:ows_general3}. Here, $\theta_{obs}=\pi/2$. Corresponding values of $\xi$ and $\eta$ can be calculated from Eq. (\ref{eq:celestial}).}
\label{fig:ows_pot}
\end{figure}

Figure \ref{fig:ows_pot1} shows the plots of the effective potential for the parameter values denoted by the black dots shown in Fig. \ref{fig:ows_genera2}. It is clear that, for the impact parameter values denoted by the black dot on the solid black, solid blue, and black dashed curve of Fig. \ref{fig:ows_genera2}, respectively, we have an unstable photon orbit at the throat, an unstable photon orbit outside the throat, and a stable photon orbit at the throat, which are in agreement with Fig. \ref{fig:ows_genera2}. Also, note that the critical impact parameters corresponding to the stable orbits at the throat cannot decide the boundary of a shadow since photons (with these impact parameters) from the distant source get turned away from some $l$ ($r$) where $V_{eff}=0$, before they reach the throat [see the black dashed plot of Fig. \ref{fig:ows_pot1}]. Figures \ref{fig:ows_pot2}) and \ref{fig:ows_pot3}), respectively, show the plots of the effective potential for impact parameter values lying inside and outside of the shadow boundary of Fig. \ref{fig:ows_general3}. Note that, since the effective potential does not have any turning point (i.e., $V_{eff}\neq 0$) for the impact parameter lying inside the shadow boundary, photons with these impact parameters plunge into the wormhole, pass through the throat, and go to the other side [see Fig. \ref{fig:ows_pot2}]. These photons are not received by the distant observer, thereby creating dark spots in the observer's sky. On the other hand, the effective potential always has a turning point (i.e., $V_{eff}= 0$) at some $l$ ($r$) for the impact parameter values lying outside the shadow boundary [see Fig. \ref{fig:ows_pot3}]. Photons with these impact parameters escape to the distant observer, thereby creating bright spots in the observer's sky. This explains why the correct shape of the shadow is given by the common region enclosed by the blue and black curves of Fig. \ref{fig:ows_genera2}.

\begin{figure}[ht]
\centering
\subfigure[$J/M^2=0.01$]{\includegraphics[scale=0.5]{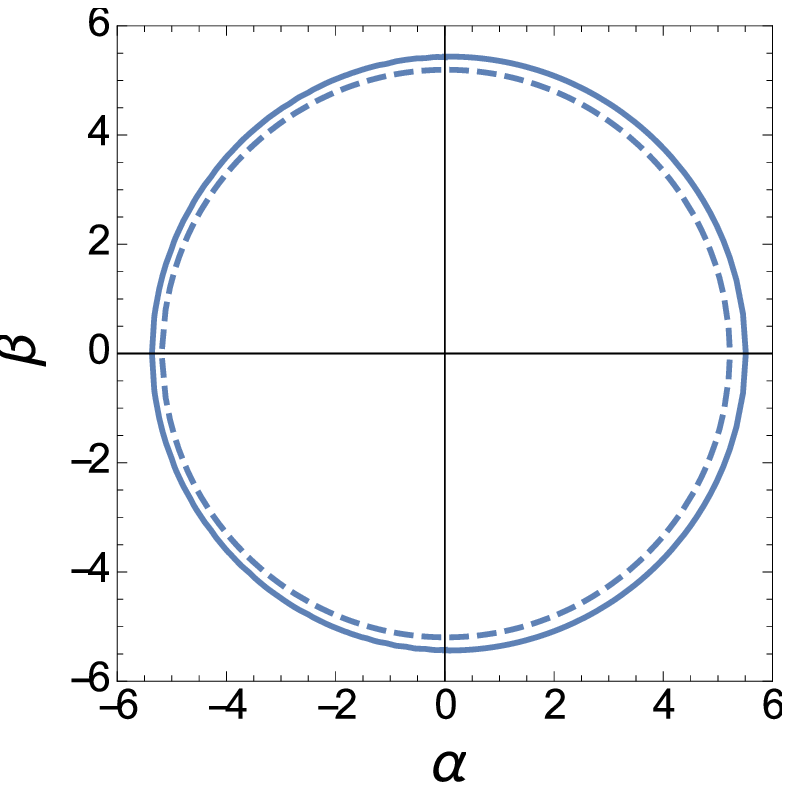}}
\subfigure[$J/M^2=0.05$]{\includegraphics[scale=0.5]{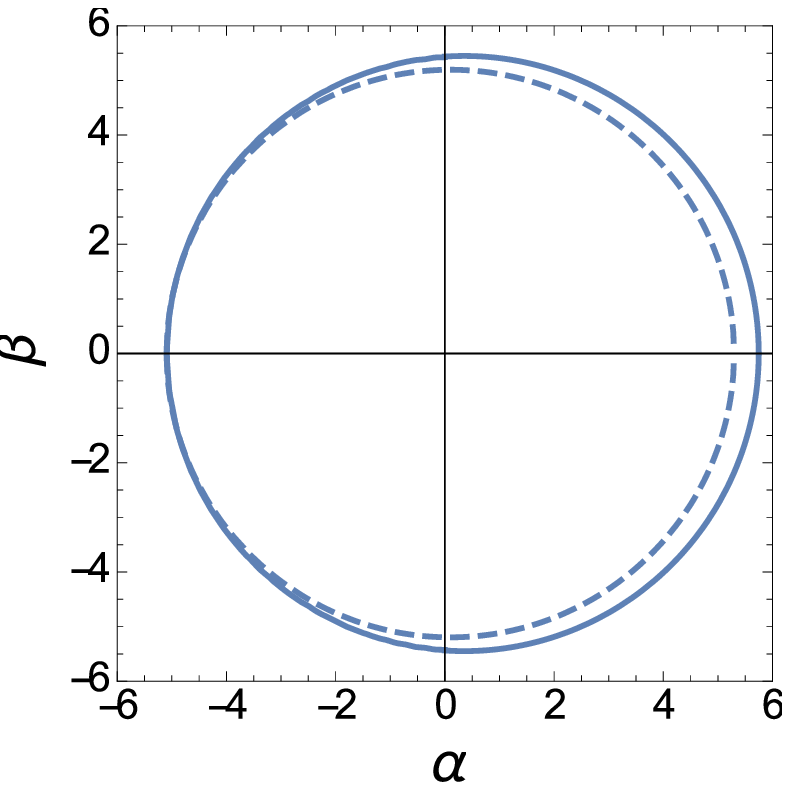}}
\subfigure[$J/M^2=0.3$]{\includegraphics[scale=0.5]{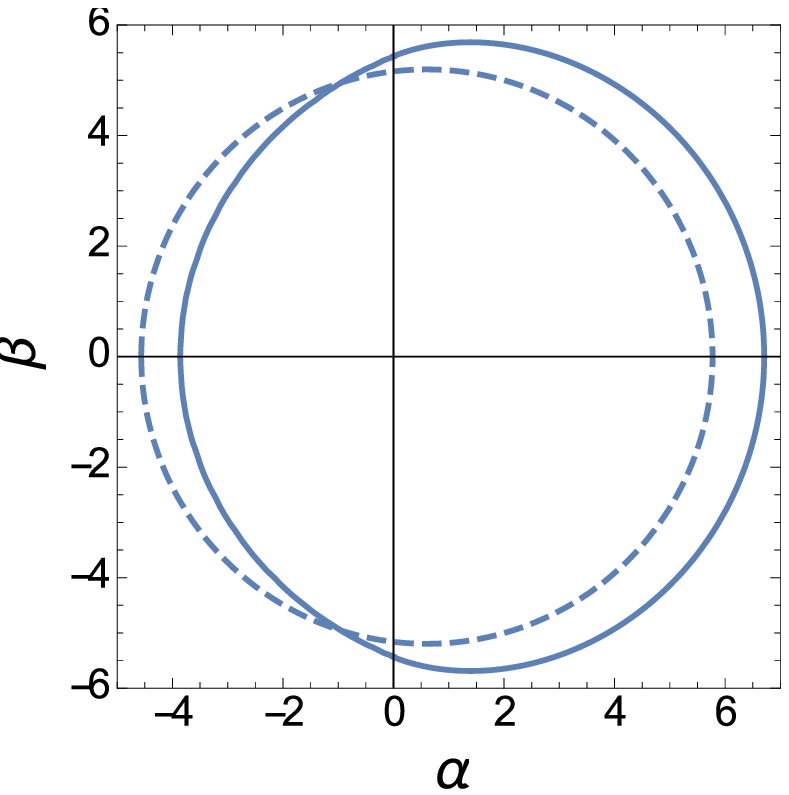}}
\subfigure[$J/M^2=0.6$]{\includegraphics[scale=0.5]{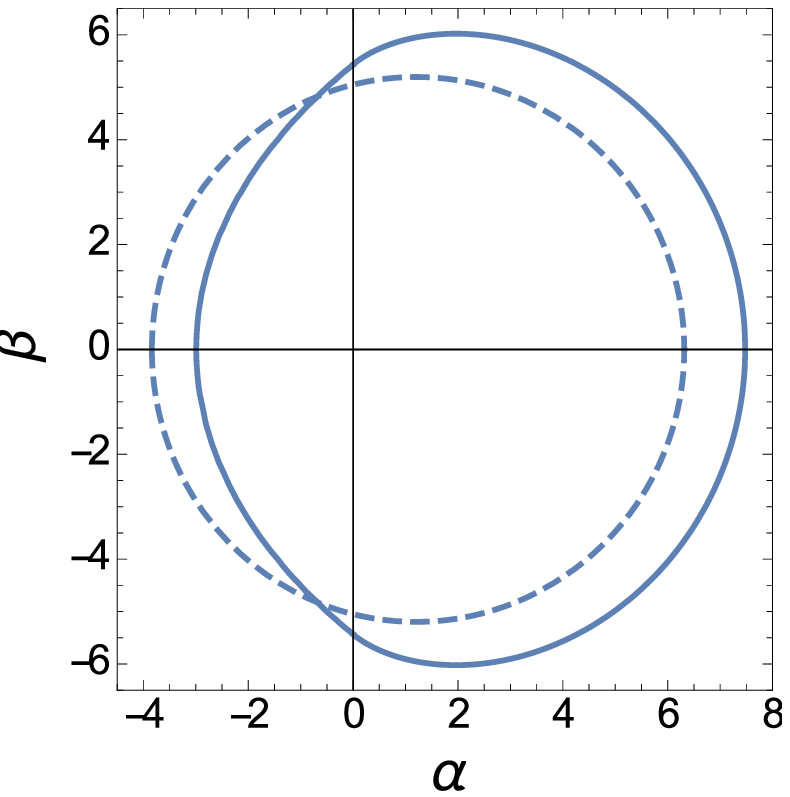}}
\subfigure[$J/M^2=0.9$]{\includegraphics[scale=0.5]{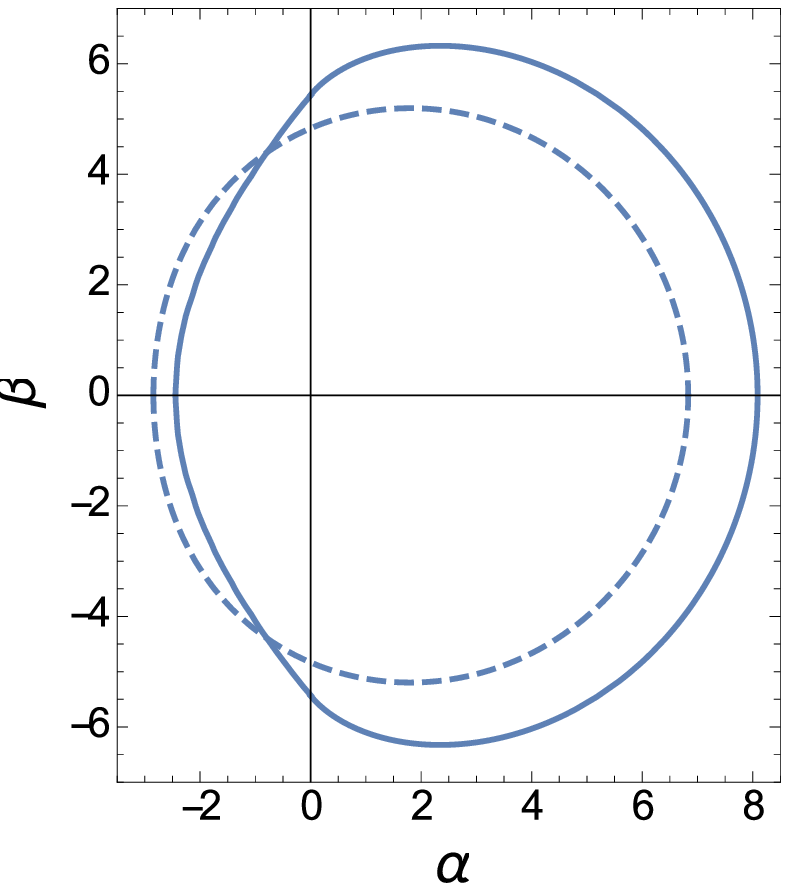}}
\subfigure[$J/M^2=1.0$]{\includegraphics[scale=0.5]{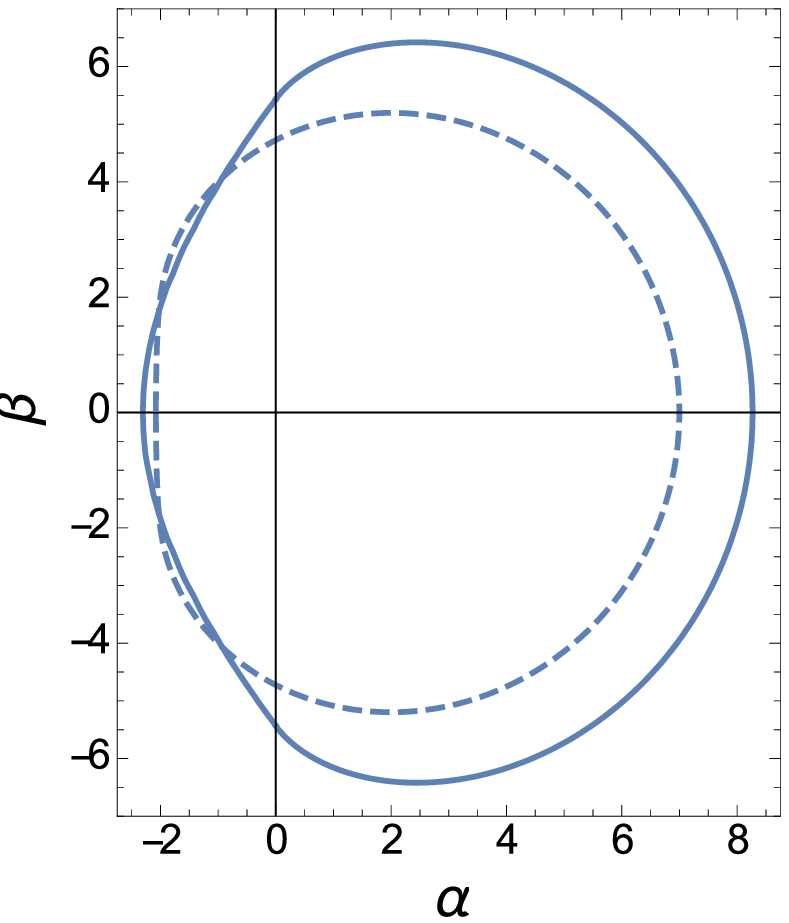}}
\subfigure[$J/M^2=1.2$]{\includegraphics[scale=0.5]{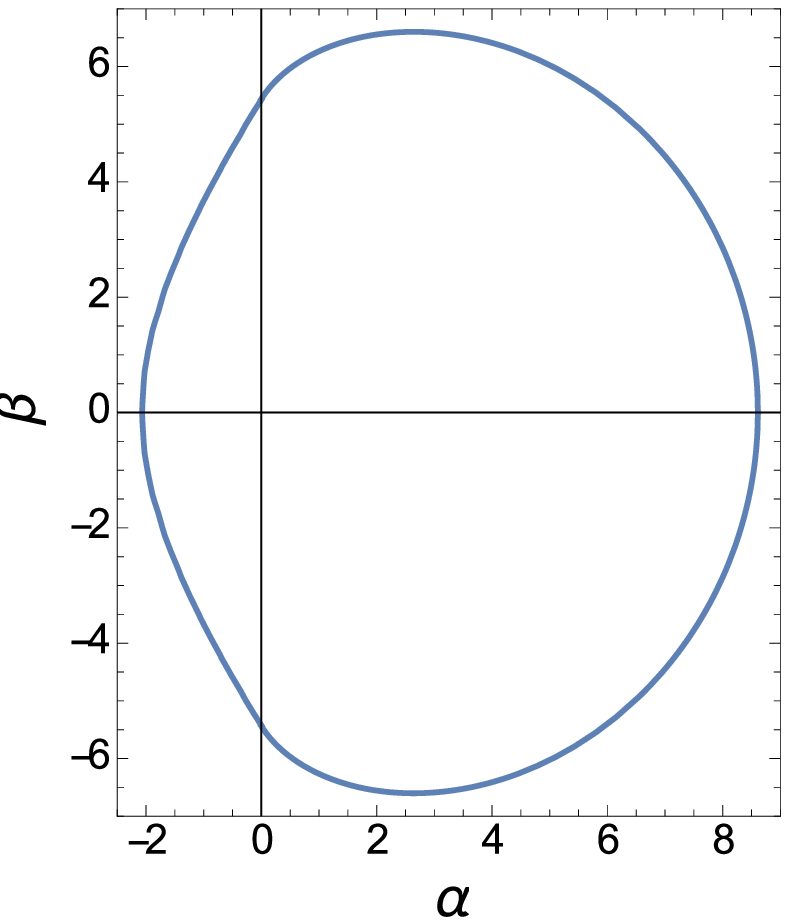}}
\subfigure[$J/M^2=2.0$]{\includegraphics[scale=0.5]{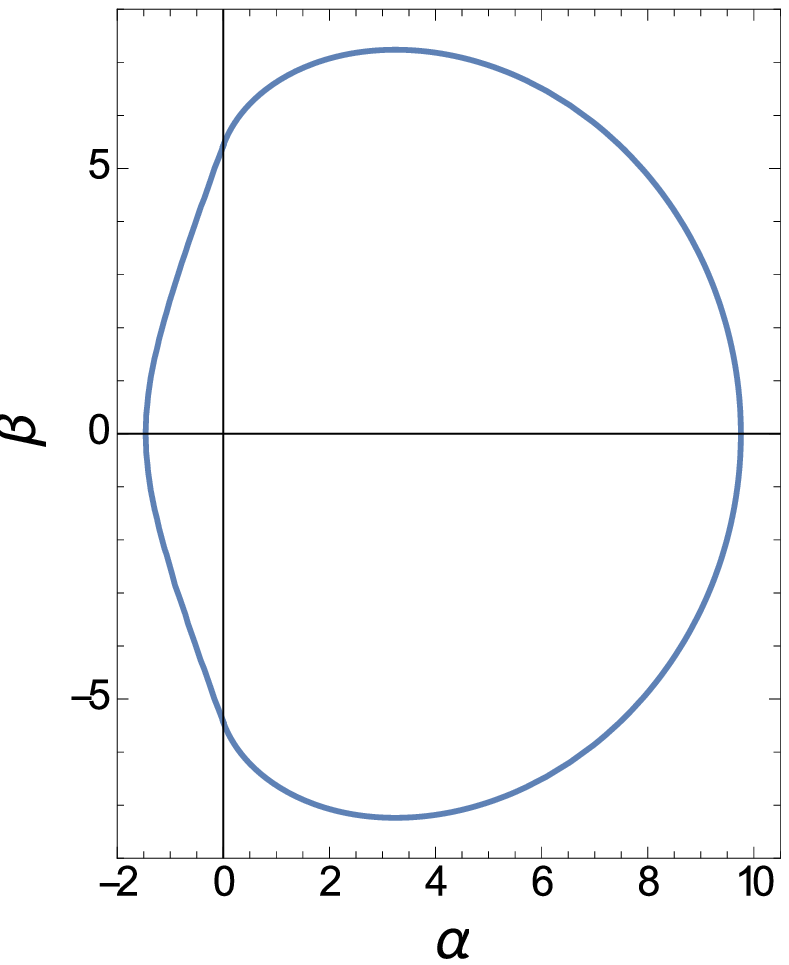}}
\caption{The shadow of a rotating wormhole (solid curve) whose metric functions are given by (\ref{eq:metric_choice_1}) and the Kerr black hole (dashed curve) for different spin values [(a)-(h)]. Here, $\theta_{obs}=90^{\circ}$. The axes are in units of the mass $M$.}
\label{fig:shadow1_90d}
\end{figure}

\begin{figure}[ht]
\centering
\subfigure[$J/M^2=0.01$]{\includegraphics[scale=0.5]{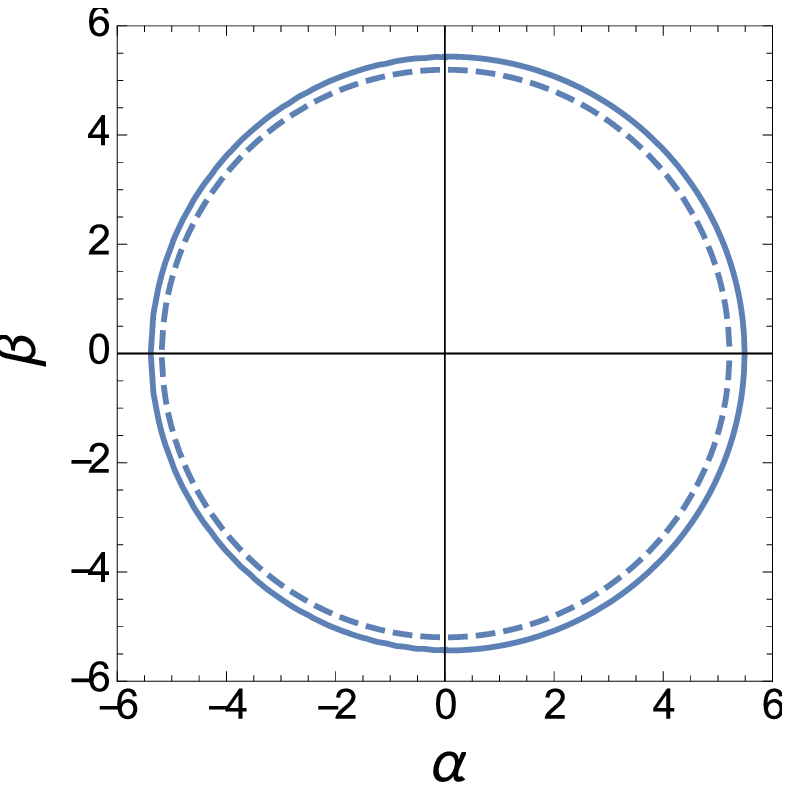}}
\subfigure[$J/M^2=0.05$]{\includegraphics[scale=0.5]{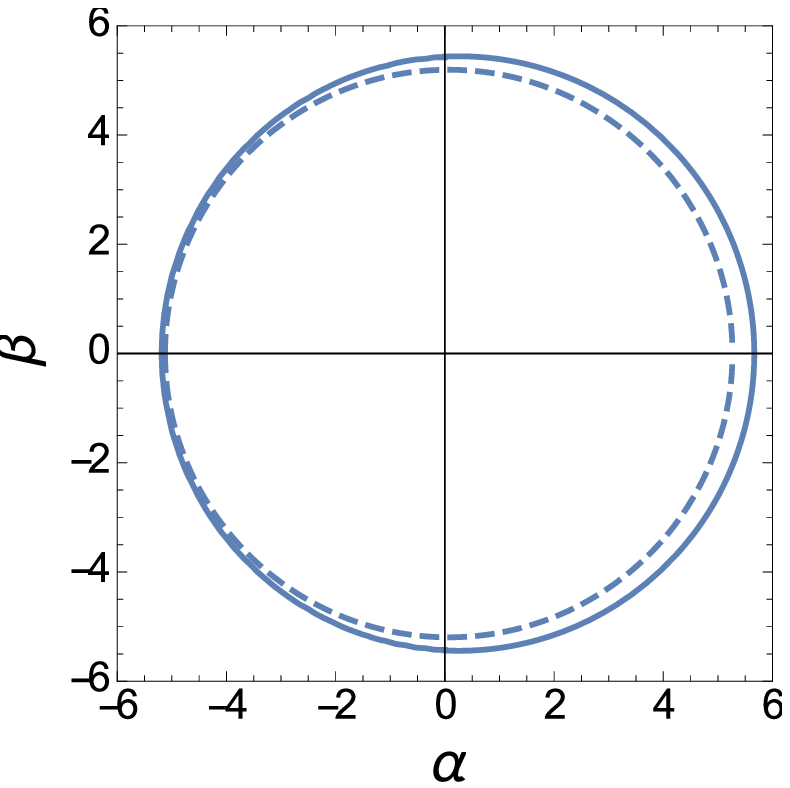}}
\subfigure[$J/M^2=0.3$]{\includegraphics[scale=0.5]{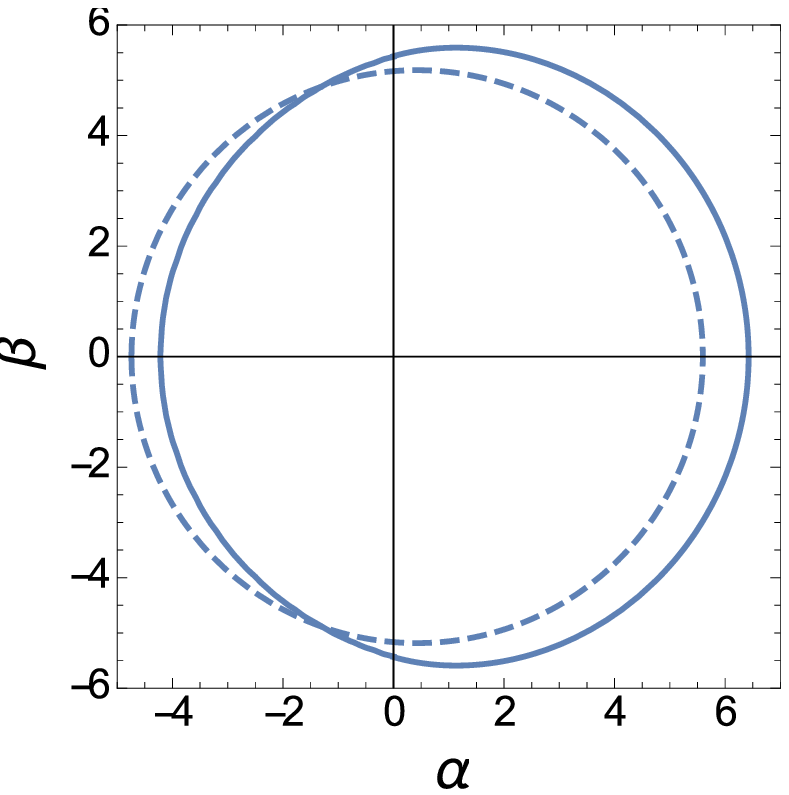}}
\subfigure[$J/M^2=0.6$]{\includegraphics[scale=0.5]{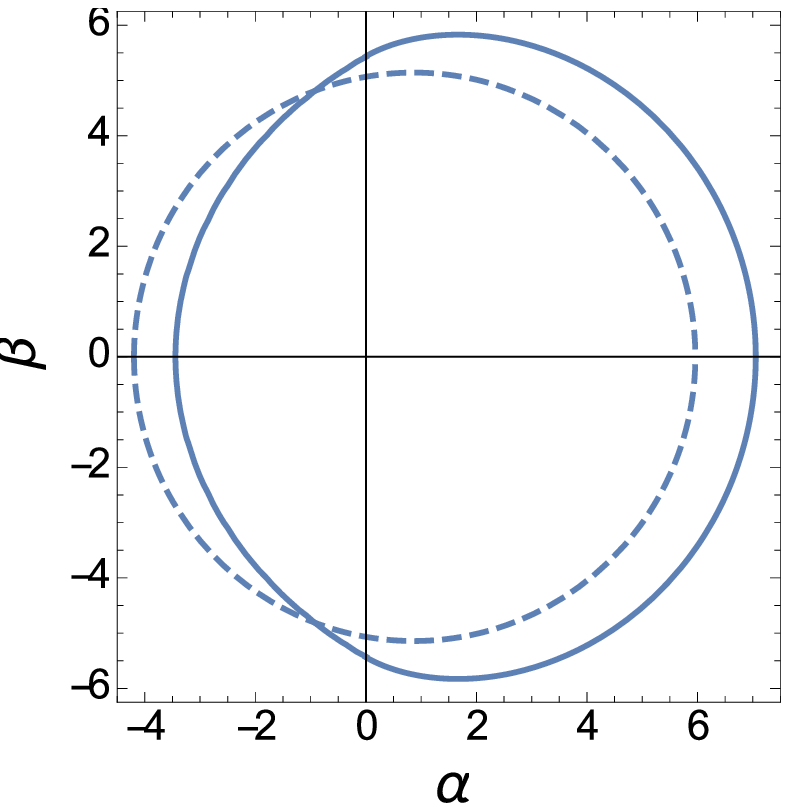}}
\subfigure[$J/M^2=0.9$]{\includegraphics[scale=0.5]{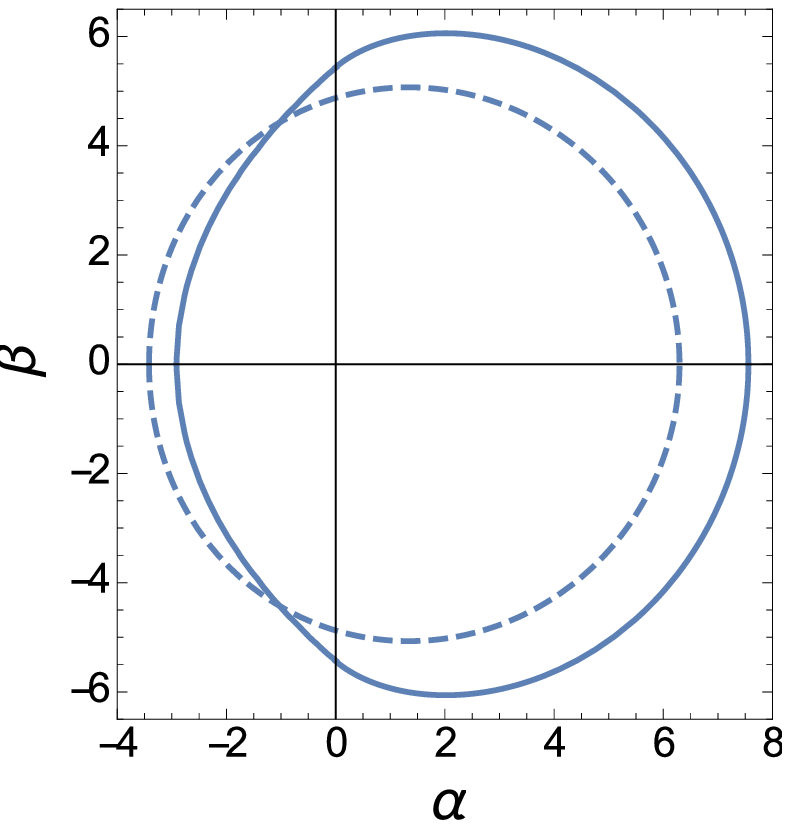}}
\subfigure[$J/M^2=1.0$]{\includegraphics[scale=0.5]{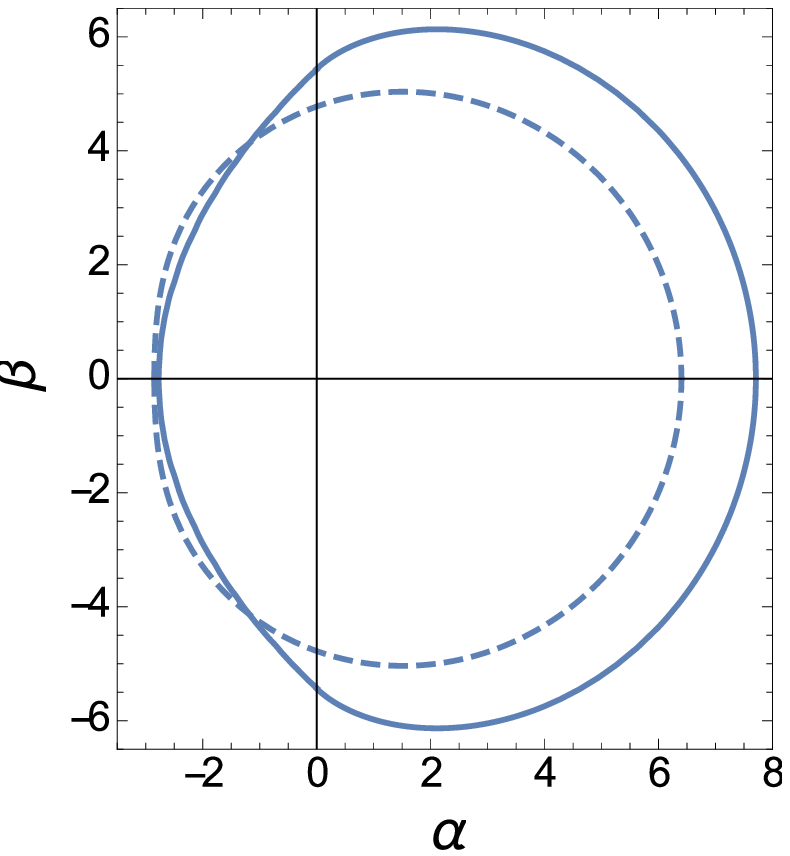}}
\subfigure[$J/M^2=1.2$]{\includegraphics[scale=0.5]{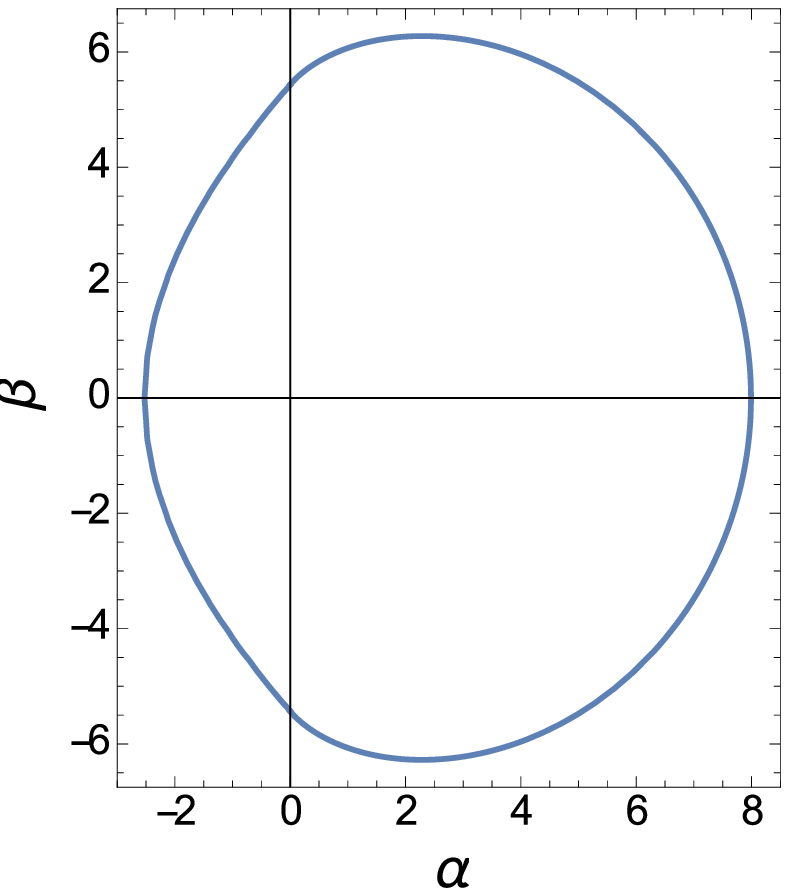}}
\subfigure[$J/M^2=2.0$]{\includegraphics[scale=0.5]{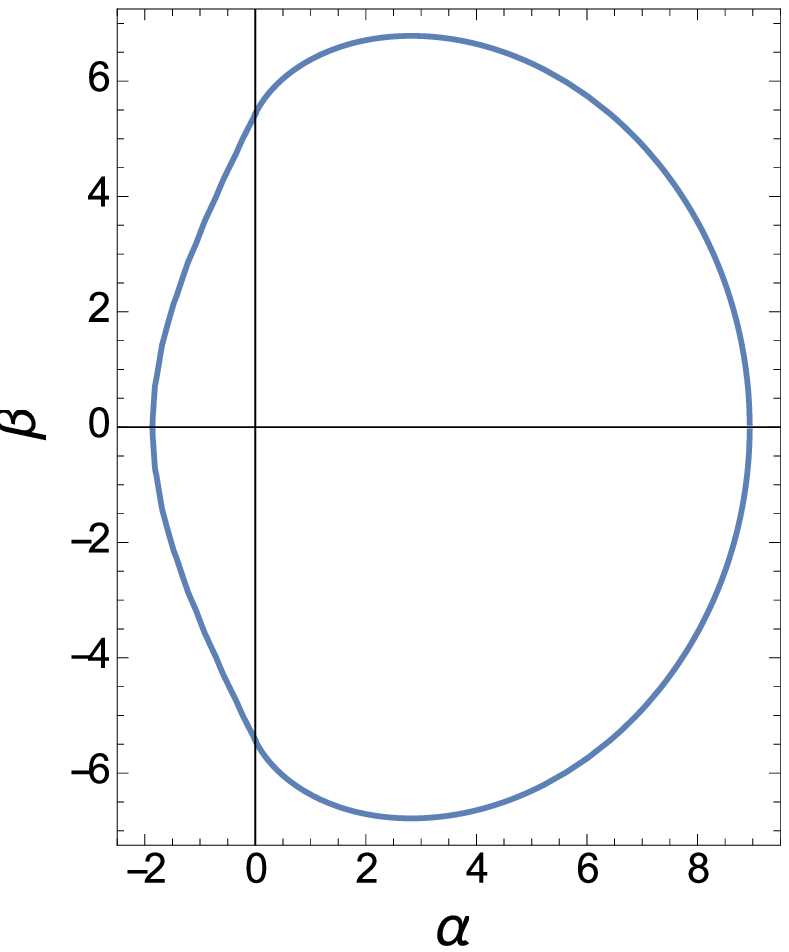}}
\caption{The shadow of a rotating wormhole (solid curve) whose metric functions are given by (\ref{eq:metric_choice_1}) and the Kerr black hole (dashed curve) for different spin values [(a)-(h)]. Here, $\theta_{obs}=45^{\circ}$. The axes are in units of the mass $M$.}
\label{fig:shadow1_45d}
\end{figure}
With the above discussions in mind, we now obtain the apparent shape of the shadow cast by a rotating wormhole. Figures (\ref{fig:shadow1_90d}) and (\ref{fig:shadow1_45d}) show the shadow cast by a rotating wormhole whose metric functions are given by (\ref{eq:metric_choice_1}). For each set of the parameter values, we have also plotted the shadow cast by a Kerr black hole for comparison. Note that, for small spin $a$ ($=J/M^2$), the wormhole shadow is qualitatively similar to that of the Kerr black hole, i.e., they both are almost circular. However, with increasing values of the spin, the characteristic deformation of the shadow (i.e., deviation from the circular shape) due to the spin is more and more prominent in the wormhole case than that in the black hole case. Such considerable deviation of the wormhole shadow from that of the Kerr black hole may be relevant to discriminate between the wormhole and the black hole in future observations.

\begin{figure}[ht]
\centering
\subfigure[$J/M^2=0.01$]{\includegraphics[scale=0.5]{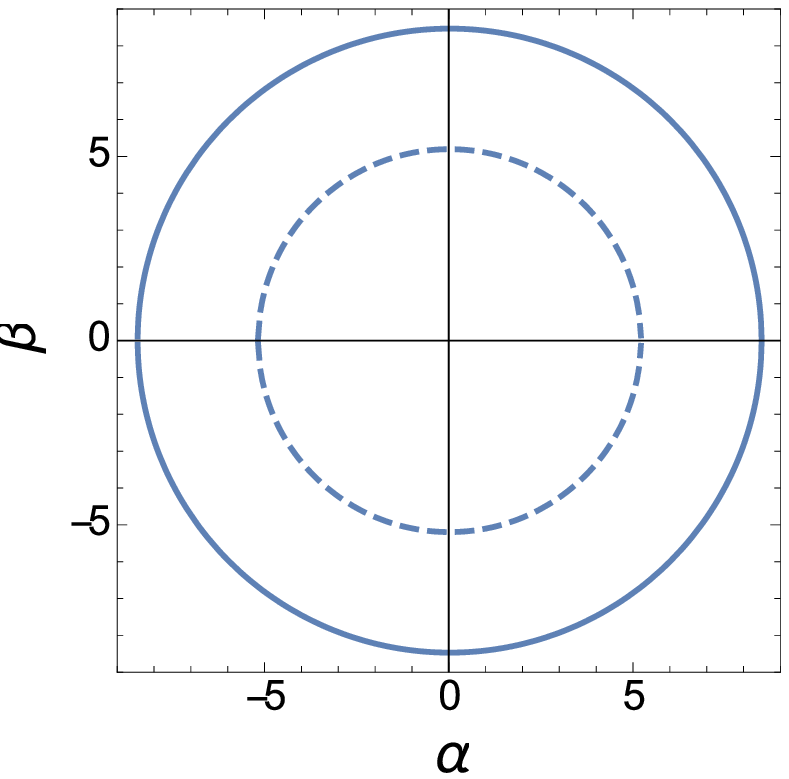}}
\subfigure[$J/M^2=0.1$]{\includegraphics[scale=0.5]{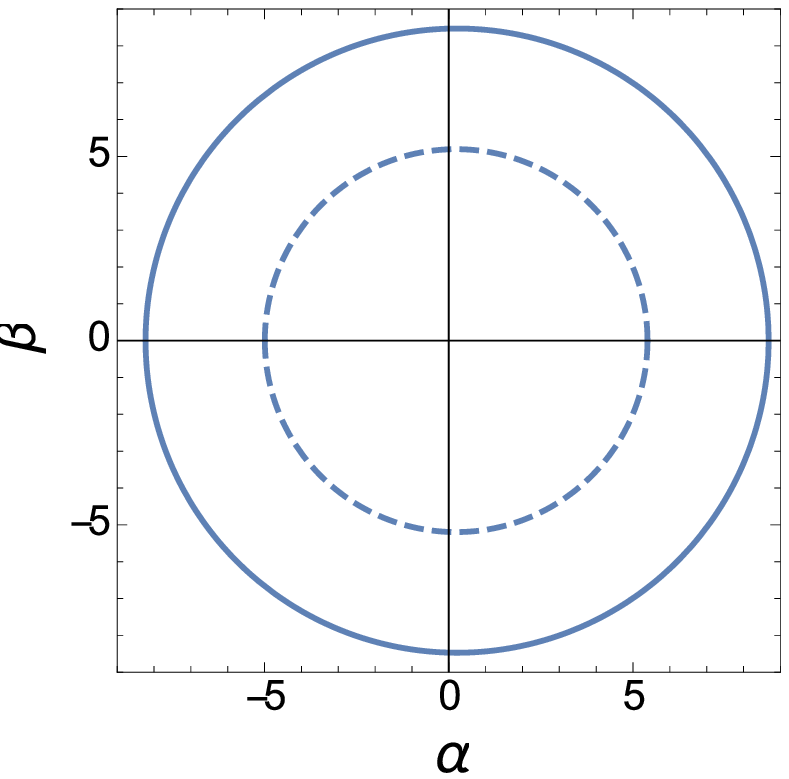}}
\subfigure[$J/M^2=0.3$]{\includegraphics[scale=0.5]{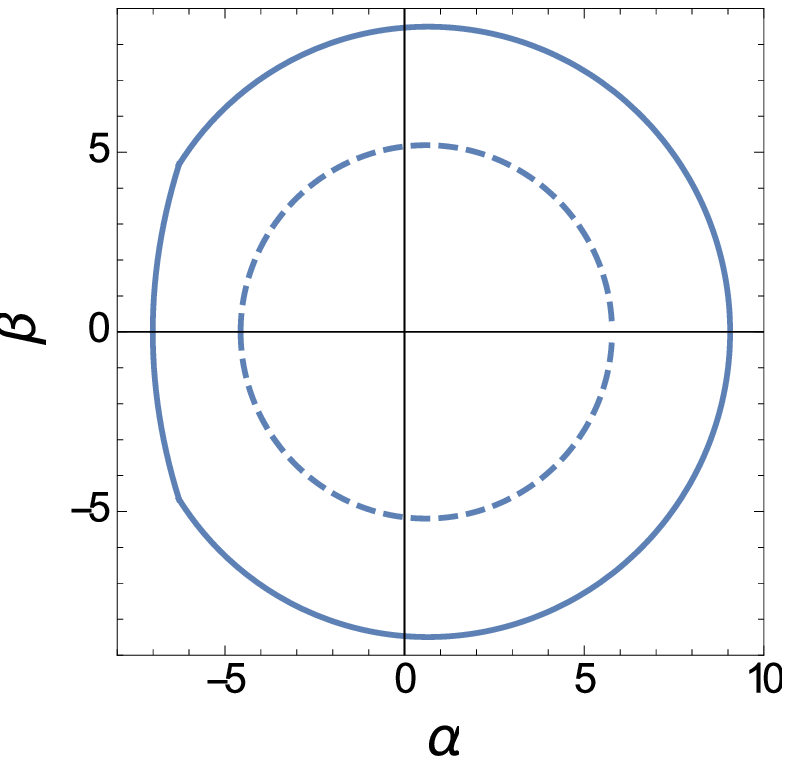}}
\subfigure[$J/M^2=0.6$]{\includegraphics[scale=0.5]{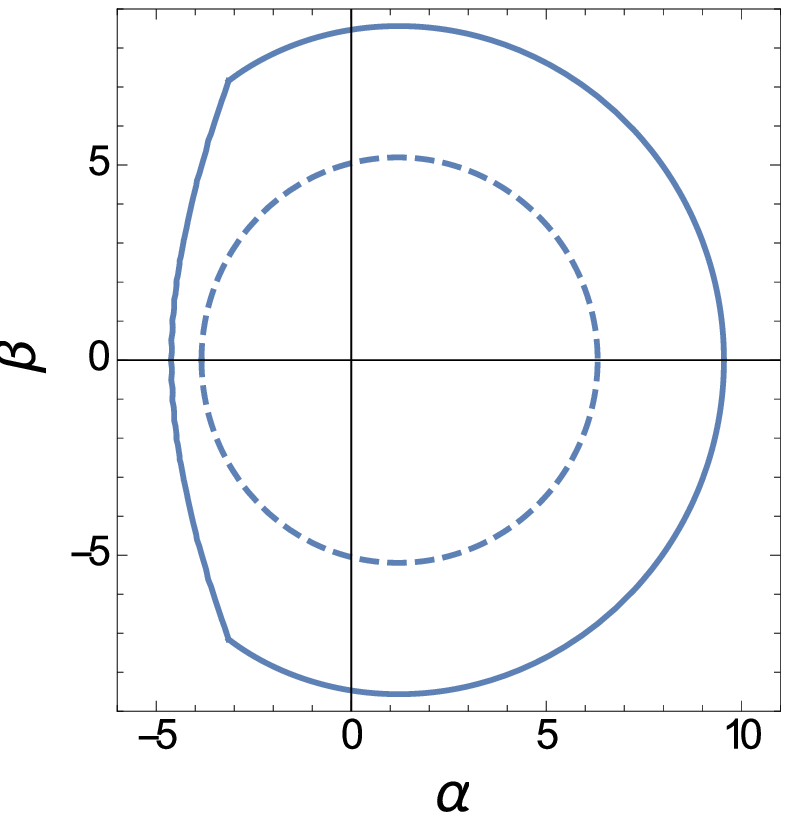}}
\subfigure[$J/M^2=0.9$]{\includegraphics[scale=0.5]{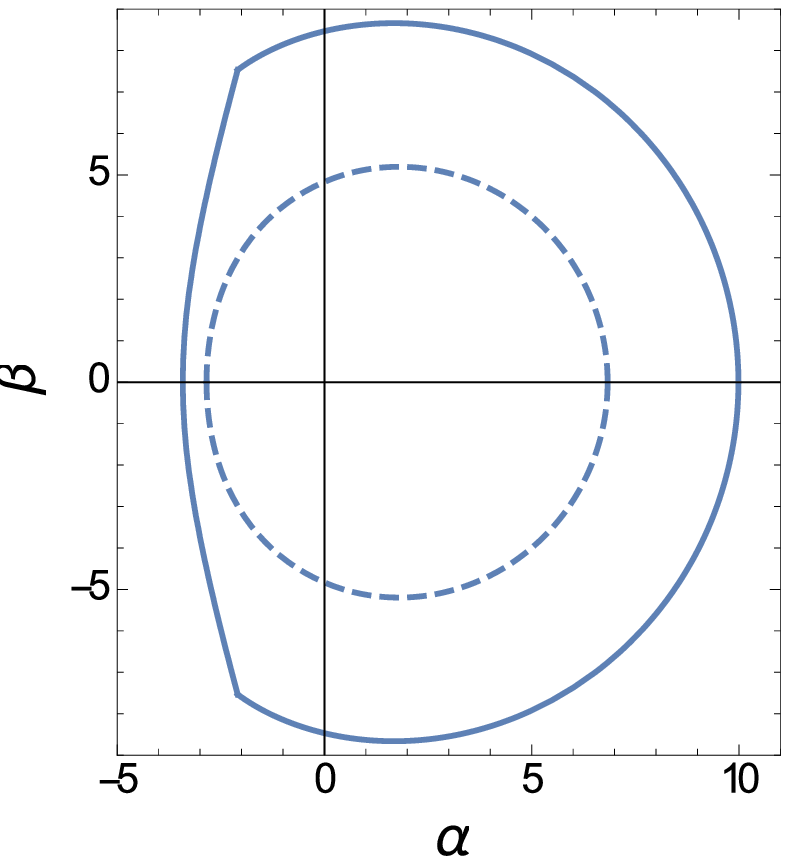}}
\subfigure[$J/M^2=1.0$]{\includegraphics[scale=0.5]{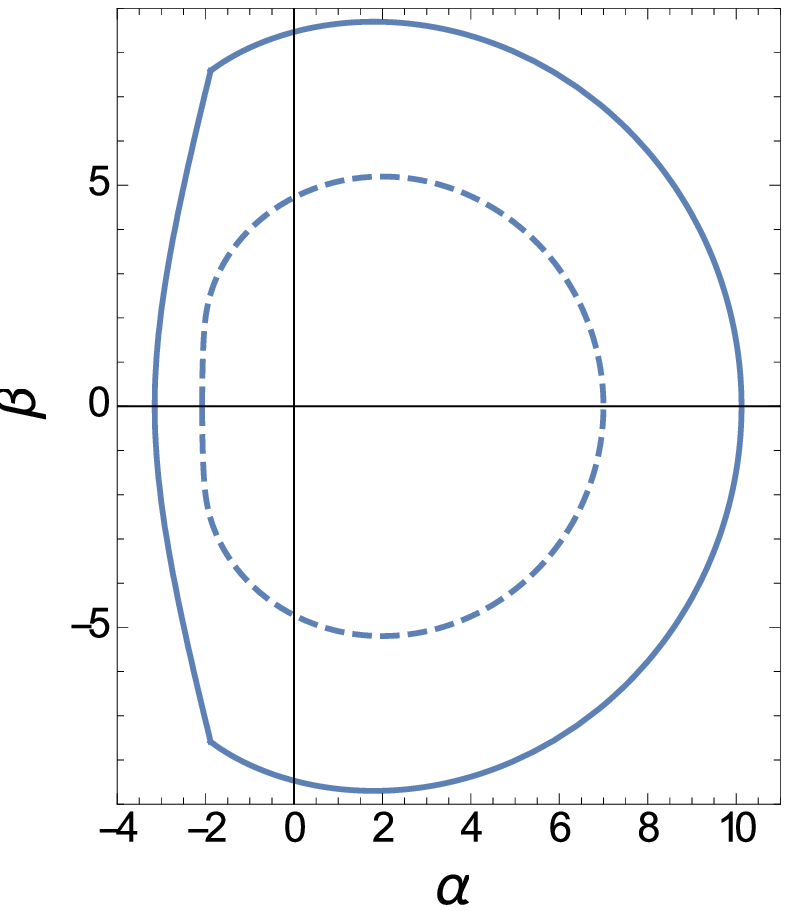}}
\subfigure[$J/M^2=1.2$]{\includegraphics[scale=0.5]{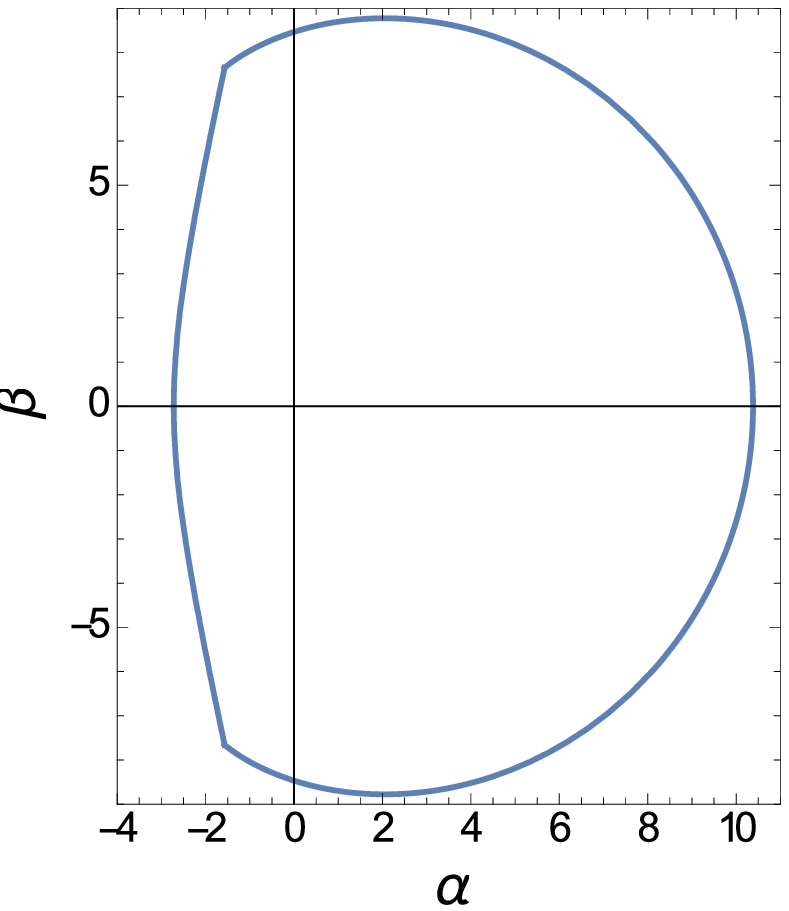}}
\subfigure[$J/M^2=2.0$]{\includegraphics[scale=0.5]{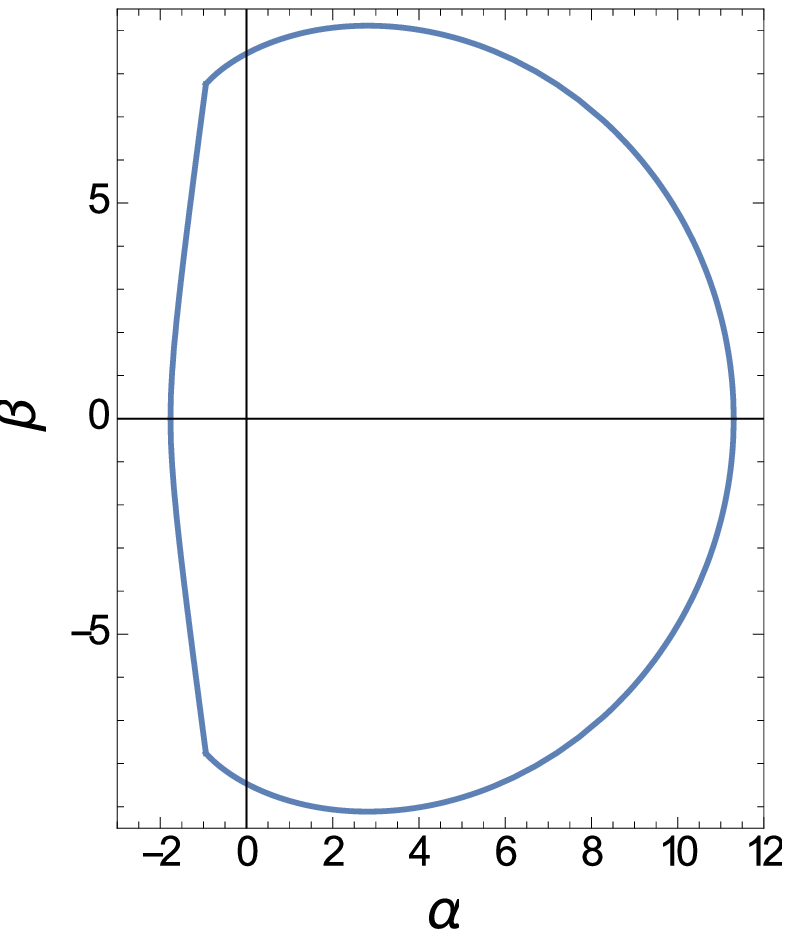}}
\caption{The shadow of a rotating wormhole (solid curve) whose metric functions are given by (\ref{eq:metric_choice_2}) and the Kerr black hole (dashed curve) for different spin values [(a)-(h)]. Here, $\theta_{obs}=90^{\circ}$. The axes are in units of the mass $M$.}
\label{fig:shadow2_90d}
\end{figure}

As a second example, we consider a wormhole whose metric functions are given by
\begin{equation}
N =\exp\left[-\frac{r_0}{r}-\frac{r_0^2}{r^2}\right], \quad b(r)=r_0=2M,\quad K=1, \quad \omega = \frac{2J}{r^3}.
\label{eq:metric_choice_2}
\end{equation}
The shadow for the above choice of the metric functions, along with that of the Kerr black hole, is shown in Fig. \ref{fig:shadow2_90d} for different values of the spin parameter. In this case, the shadow of the wormhole is larger than that of the black hole for all spins. Also, note that, similar to the earlier case, in this case too, the wormhole shadow starts deviating from the black hole one as we increase the spin.

\section{Conclusion}
\label{sec:conclusion}
In this work, we have studied shadows cast by rotating wormholes. We have discussed in detail why and how the throat of a rotating wormhole plays a crucial role in the shadow formation. This crucial role of the throat has been overlooked in the previous studies \citep{nedkova_2013,abdujabbarov_2016b} on the shadow of the same class of rotating wormholes considered here. Therefore, the results obtained in the above-mentioned earlier works are incomplete. We have revisited the problem and obtained the correct shapes of the shadows. We have compared our results with that of the Kerr black hole. For small spin, the wormhole shadow is qualitatively similar to that of the Kerr black hole, i.e., they both are almost circular. However, with increasing values of the spin, the shapes of the wormhole shadows start deviating considerably from that of the black hole. Such considerable deviation, if detected in future observations, may possibly indicate the presence of a wormhole. In other words, the results obtained here indicate that, through the observations of their shadows, a wormhole having reasonable spin, can be distinguished from a black hole.

However, our conclusions are largely based on the types of wormholes we have chosen to work with. 
It will be interesting to see whether or to what extent the conclusions drawn here carry over to a broader class of rotating wormholes.

\section*{Acknowledgments}
\noindent The author thanks Professor Sayan Kar for going through the ADM mass calculation. He also thanks the anonymous referee for making critical comments which helped to improve the manuscript.

\end{document}